\documentclass{article}
\usepackage{caption}
\usepackage{subcaption}
\usepackage{hyperref}
\usepackage{authblk}
\usepackage{cite}
\usepackage{amsmath,amssymb,amsfonts}
\usepackage{algorithm}
\usepackage{algorithmic}
\usepackage{multirow}

\usepackage{graphicx}
\usepackage{textcomp}

\title{The Adversarial Security Mitigations of mmWave Beamforming Prediction Models using Defensive Distillation and Adversarial Retraining }

\author[$\dagger$]{Murat Kuzlu}
\author[2]{Ferhat Ozgur Catak}
\author[3]{Umit Cali}
\author[4]{Evren Catak}
\author[5]{Ozgur Guler}

\affil[$\dagger$]{Department of Engineering Technology, Old Dominion University, Norfolk, VA, USA, (e-mail: mkuzlu@odu.edu)}
\affil[$2$]{Department of Electrical Engineering \& Computer Science, University of Stavanger, Rogaland, Norway (e-mail: f.ozgur.catak@uis.no)}
\affil[$3$]{Department of Electric Power Engineering, Norwegian University of Science and Technology, Trondheim, Norway (e-mail: umit.cali@ntnu.no)}
\affil[$4$]{Independent Researcher, Stavanger, Norway (e-mail: evren.catak@ieee.org)}
\affil[$5$]{eKare Inc., Fairfax, VA, USA, (e-mail: oguler@ekare.ai)}

\date{}

\begin{document}

\maketitle

\begin{abstract}
The design of a security scheme for beamforming prediction is critical for next-generation wireless networks (5G, 6G, and beyond). However, there is no consensus about protecting the beamforming prediction using deep learning algorithms in these networks. This paper presents the security vulnerabilities in deep learning for beamforming prediction using deep neural networks (DNNs) in 6G wireless networks, which treats the beamforming prediction as a multi-output regression problem. It is indicated that the initial DNN model is vulnerable against adversarial attacks, such as Fast Gradient Sign Method (FGSM), Basic Iterative Method (BIM), Projected Gradient Descent (PGD), and Momentum Iterative Method (MIM), because the initial DNN model is sensitive to the perturbations of the adversarial samples of the training data. This study also offers two mitigation methods, such as adversarial training and defensive distillation, for adversarial attacks against artificial intelligence (AI)-based models used in the millimeter-wave (mmWave) beamforming prediction. Furthermore, the proposed scheme can be used in situations where the data are corrupted due to  the adversarial examples in the training data. Experimental results show that the proposed methods effectively defend the DNN models against adversarial attacks in next-generation wireless networks.
\end{abstract}

\section{Introduction}
\subsection{Preamble}
The first 5G standard was announced and approved by 3GPP in December 2017 \cite{8403769}. The early standardization work on 5G is expected to provide a solid and stable foundation for the early adoption of 5G services. In addition, 5G will be essential  for Internet of Things (IoT) applications and future mobile networks. There are many challenges in the design of 5G networks \cite{CATAK2017184}, including a security scheme for beamforming prediction. It is a critical part of wireless networks, which has been studied in communication systems and signal processing. It is also crucial to design and implement beamforming algorithms in next generation (i.e., 6G) wireless networks. In current wireless networks, deep learning (DL)-based beamforming prediction is vulnerable to adversarial machine learning attacks \cite{catak2021security}. Therefore, it is critical to design a security scheme for beamforming prediction in 6G networks. 

6G is the latest wireless communication technology among cellular networks that is currently under development. In 6G solutions, artificial intelligence (AI)-based algorithms, especially DL, would be one of the main components of wireless communication systems \cite{zheng2021potential} to improve the overall system performance. The existing solutions in 5G would be migrated to the AI domain, more specifically into the DL area. Therefore, it is crucial to design secure DL solutions for the AI-based models in 6G wireless networks. The new attack surface, in addition to the existing 5G security problems, is DL security vulnerabilities. Researchers and companies should mitigate their DL models' security problems before deploying them to the production environments. They need to identify, document, and perform risk assessment for the new types of security threats in the next-generation wireless communication systems.

\subsection{Related Works}
5G networks were commercially launched in late 2018 \cite{liu20185g}. After the first commercial launch of the 5G network, the planning of the next generation networks, such as 6G, commenced to provide communication services for future demands. The most important key for this next generation is the use of advanced communications and AI
technologies \cite{de2021survey}. In the literature, there are many studies focused on next-generation wireless networks (5G, 6G, and beyond) and the integration of current emerging AI tools into these networks \cite{zhang20196g,giordani2020toward,saad2019vision,9163104, sheth2020taxonomy}. Next-generation wireless networks have been considered as one of the most important drivers in in the ability of current and future information age applications (i.e., virtual and augmented reality, remote surgery, holographic projection, metaverse, etc.) to meet the forecast requirements, such as ultra-broadband, ultra-reliable, low latency communication, massive access, and real-time services with low cost.  The authors in \cite{9206115} reviewed AI-based solutions in 6G networks to achieve these requirements, and emphasized several solutions for an ultra-broadband transmission (terahertz channel estimation and spectrum management), secure communication (authentication, access control, and attack detection), and ultra-reliability and low latency services (intelligent resource allocation). The study \cite{9023459} investigated next-generation wireless networks in terms of core services, key performance indices (KPIs), enabling technologies, architecture, challenges and possible solutions, opportunities, and future research trends. It also evaluated core services for 5G and  6G networks. Further, it also indicated that several emerging technologies will play a key role in 6G networks, i.e.,  AI for improving the system performance, blockchain for managing the system security and quantum computing for computing efficiency. The authors of \cite{ozpoyraz2022deep} provided a comprehensive review of DL-based solutions focusing on emerging physical layer techniques such as massive multiple-input multiple-output (MIMO), multi-carrier (MC) waveform, reconfigurable intelligent surface (RIS) communications, and security, for 6G networks. It also indicated that AI will significantly contribute to next-generation networks to improve their performance. The study \cite{ali20206g} addressed the key role of next-generation networks for humans and systems and discussed how ML-based solutions will improve these networks in terms of performance, control, and security, and  solve problems in various network layers, i.e., the physical, medium access, and application. Many researchers of 6G networks have explored AI by adopting it as the top solution in many extremely complex scenarios. Yang et al. \cite{9237460} presented an AI-enabled intelligent  6G networks architecture which can support several services such as discovery,  automatic network adjustment, smart service provisioning , and intelligent resource management. It also discusses AI-based methods and how to apply them to 6G networks by efficiently optimizing network performance, including intelligent spectrum management, mobile edge computing, mobility and handover management. 

\subsection{Purpose and Contributions}
Utilizing DL-based algorithms for the next-generation wireless network is a great opportunity to be able to improve the overall system performance. However, it may lead to potential security problems, i.e., AI-model poisoning. While AI-based algorithms offer significant advantages for 6G networks, potential security issues related to AI-based models are typically overlooked. As such, the wireless research community should give particular attention to the security and privacy concerns regarding next-generation networks \cite{dang2020should}. The authors of \cite{porambage20216g} provide an overview of the 6G wireless networks in terms of the security and privacy challenges along with promising security solutions and technologies together with 6G networks specifications. The authors of  \cite{kuzlu2021role} investigate the role of AI in IoT security for possible cyber attacks, emphasizing the model poisoning attack, i.e., where a machine learning model's training data are poisoned.
The study \cite{siriwardhana2021ai} provides a comprehensive review of the opportunities and challenges in AI-based security and private provision as well as proposes solutions for 6G and beyond networks.

In our recent works \cite{catak2021security} and \cite{9527756}, we only investigated FGSM attacks, which can be mitigated using the adversarial training method. In this work, four different adversarial machine learning methods (FGSM, BIM, PGD, and MIM) are investigated to build robust beamforming DL models using two mitigation methods. The DL-based beamforming prediction solutions provide satisfactory results; however, these solutions cannot work under an attack, such as adversarial machine learning attacks. This paper presents a DL security scheme for beamforming prediction using deep neural networks (DNNs) in 6G wireless networks, which treats the beamforming prediction as a multi-output regression problem. The results showed that the proposed scheme is more secure against adversarial attacks because it is robust to the perturbations of the adversarial samples of the training data. 

\subsection{Organization}
The rest of the paper is organized as follows: Section \ref{sec:cybsec-frameworks} describes two publicly available cyber attack tools and the proposed framework. Section \ref{sec:background} covers the background information regarding adversarial machine learning and mitigation methods. Section \ref{sec:overview} and \ref{sec:experiments} describe the system overview, and experiments for three scenarios, respectively. Section \ref{sec:discussion} discusses the
proposed scheme along with observations, and Section  \ref{sec:conclusion}  concludes the paper.

\section{Cybersecurity Frameworks}\label{sec:cybsec-frameworks}
Cybersecurity frameworks help enterprises manage potential cyber risks in a better way and decide future plans for any cyber threat detection, or investigating a security incident in application and system development. Widely used cybersecurity frameworks are discussed along with the proposed framework below.

\subsection{ML Cyber Kill Chain}
Lockheed Martin's \textit{Cyber Kill Chain} is a methodology designed to help companies assess the risks they face and the potential impact on their organization.\footnote{https://www.lockheedmartin.com/en-us/capabilities/cyber/cyber-kill-chain.html} The methodology breaks down the seven phases of a cyber-attack and the critical activities performed during each step. The seven phases are 1. \textit{Reconnaissance}  2. \textit{Weaponization} 3. \textit{Delivery}  4. \textit{Exploitation}  5. \textit{Installation}  6. \textit{Command and Control/Actuation}, and  7. \textit{Actions on Objectives}.
By assessing the activities that take place during each phase of their organization’s potential cyber-attack, users can understand the impact of a successful cyber-attack on their organization. For the 6G wireless networks, such an assessment can help users understand the potential impact of a cyber-attack on their AI-based wireless communication solutions and their ability to continue to operate your wireless network. 
To conduct such an assessment, users first need to create a table that lists each phase of the cyber kill chain and the possible activities that could take place during each phase.
 
\subsection{MITRE ATT\&CK and Atlas}
MITRE ATT\&CK is a framework designed to enable analysts and defenders to identify the stages of an attack and construct and execute a response plan.\footnote{https://attack.mitre.org} MITRE ATT\&CK is a short, descriptive name given to each of the different stages of a cyber-attack. These are not necessarily sequential but instead are a set of high-level steps that an adversary is likely to use to accomplish a goal. The goal is to provide a common language and framework that organizations can use to describe their security processes and communicate. This makes attackers' different techniques and tactics more identifiable and easier to track. MITRE ATT\&CK is a comprehensive catalog of attack techniques used by both state and non-state actors. It allows organizations to track a potential adversary's movement and understand their methods to gain access and move laterally across a network. It is more than a list of attack vectors; it is a catalog of adversary tradecraft and behaviors that can be used to identify malicious attackers' activity and generate a more effective response strategy. This framework was designed to be used as a common language and modular so that organizations can determine on which techniques they need to focus.

MITRE developed another framework for AI-based applications, namely MITRE Atlas (Adversarial Threat Landscape for Artificial-Intelligence Systems).\footnote{https://atlas.mitre.org} It is a knowledge resource for AI systems that includes adversary tactics, methodologies, case studies based on real-world demonstrations from security groups, the state-of-the-art from academic research. It is similar to the MITRE Att\&ck framework. 

\subsection{Proposed Framework}
\begin{figure*}[htbp]
    \centering
    \includegraphics[width=1.0\linewidth]{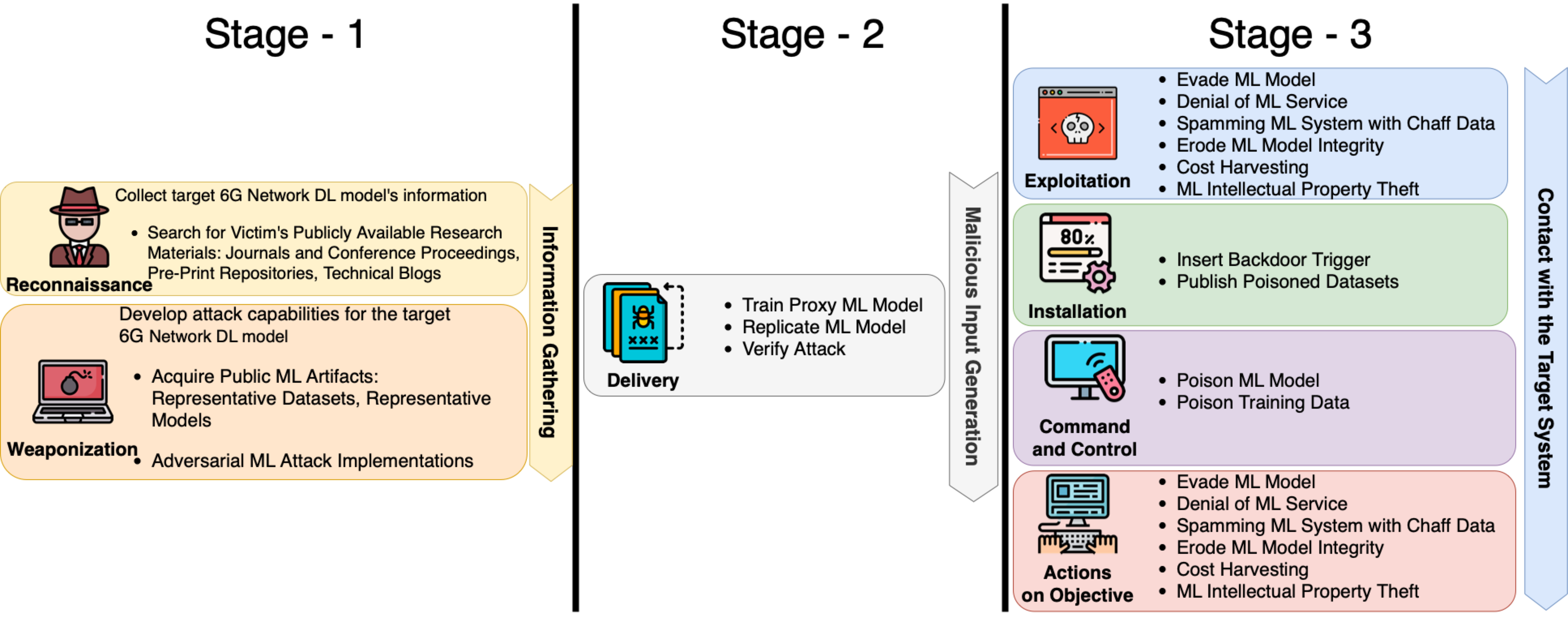}
    \caption{Cyber kill chain for AI-based applications of 6G wireless communication networks }
    \label{fig:cyber_kill_chain}
\end{figure*}

In this study, the Cyber Kill Chain and MITRE Atlas frameworks are matched to detect and fix the vulnerabilities of ML models, which will be the new component of potential AI-based 6G networks. In this way, we aim to show both threats and protection methods. The DL-based beamforming prediction models are used for MIMO systems as the proof-of-concept study of new cyber threats for 6G networks. Figure \ref{fig:cyber_kill_chain} shows the cyber kill chain for AI-based applications with 3 stages.

Stage-1 is the process of acquiring artefacts about the target AI model for beamforming prediction (i.e., reconnaissance) and building the adversarial machine learning generated craftily designed noise (i.e., weaponization). The adversary can collect the artefacts of AI models in 6G solutions (e.g., AI models in base-stations), such as the weights and hyper-parameters used in the training process and datasets from publicly available resources. After this step, the adversary can replicate the AI model in its environment to build malicious inputs. Stage-2 is the process of building the replicated model, finding the vulnerabilities and building the craftily generating malicious pilot signals (i.e., inputs of the beam prediction model) into the target AI model (i.e., delivery). Lastly, Stage-3 is the process of executing the target AI model with the malicious input signals (i.e., exploitation and installation), which can cause the AI model to produce erroneous results. The adversary can then use the malicious input signals to exploit the AI model and install the backdoor in the AI model (i.e., command and control). The adversary can use the backdoor to take control of the AI model and the target system (actions on objective).

The detailed information on the hostile tactics and methodologies parts of MITRE Atlas is given below, which will take place in the Cyber kill chain stages: i) In the reconnaissance phase, the adversary gathers information about the organization and its networks, systems, and employees. This information can be used to build a profile of the organization, the employees working there, and the organization's network and systems. This information can then be used to make a social engineering attack. ii) In the weaponization phase, the adversary uses the information gathered during the reconnaissance phase to develop the tools they need to launch an attack against the organization successfully. The adversary can then focus on the delivery phase, using the same tools to deliver information or files to the organization's network. The adversary will use the information they gathered during the reconnaissance phase to determine the best delivery mechanism to get the information it wants to deliver to the organization's network. Once the adversary has delivered the information, they need to exploit a vulnerability in the organization's network. In this phase, it can use the information gathered during the reconnaissance phase to identify the software operated by the organization, operating systems, and applications running on the organization's systems. During the exploitation phase, the adversary uses some of the information gathered in the reconnaissance phase to identify the best way to exploit the organization's network. They can use the reconnaissance phase information to identify the best software, operating systems, and applications to exploit the organization's network. When the adversary has exploited the organization's network, they have the ability to install malicious software on the organization's systems. This malicious software can then exploit the organization's network further or monitor the organization's network. During the command and control phase, the adversary can use the malicious software installed during the exploitation phase to install additional malicious software on the organization's systems. This malicious software can then be used to control the organization's systems. During the actions on objectives phase, the adversary can use the malicious software installed during the exploitation phase to access the organization's systems and steal information or interfere with the organization's network. After the adversary has completed all the steps of the cyber kill chain, they have been able to launch a successful cyberattack on the organization's network. The organization's ability to continue to operate its network can be affected by the adversary's activities during each phase of the cyber kill chain. 

\section{Background}
\label{sec:background}

In this section, a brief overview of the beamforming prediction, the existing adversarial machine learning attacks, such as Fast Gradient Sign Method (FGSM), Basic Iterative Method (BIM), Projected Gradient Descent (PGD), and Momentum Iterative Method (MIM), along with the existing solutions, i.e., adversarial training and defensive distillation, for the beamforming prediction in 6G wireless networks are presented. We also introduce the proposed scheme using deep neural networks to protect the beamforming prediction in 6G wireless networks.

\subsection{Adversarial Machine Learning}
In adversarial machine learning, the attacker tries to generate a perturbation to the adversarial examples, which would affect the prediction phase of the machine learning model \cite{2021arXiv210204150F}. The goal of the attacker is to manipulate the trained model output so that the attacker can benefit from the user's perspective. Adversarial machine learning attacks work well if the attacker has the access to the training data. However, the proposed scheme is robust to the perturbations of the adversarial samples of the training data, which in turn makes the proposed scheme robust to adversarial machine learning attacks.

Taking beamforming prediction DL model as an example, here, we use $h(\mathbf{x},\omega): \mathbb{C}^k \mapsto \mathbb{R}^m$ to denote that an uplink pilot signals, $\mathbf{x} \in \mathbb{C}^k$, to beamforming vectors, $\mathbf{y} \in \mathbb{R}^m$ where $\omega$ shows the parameters of the prediction model, $h$. Given the budget $\epsilon$ (i.e., the norm vector of the noise), the attacker tries to find a noise vector $\sigma \in \mathbb{C}^k$ to maximize the loss function $\ell$ output \cite{2021arXiv210201356B}. The attacker uses the lowest possible budget to corrupt the inputs, aiming to increase the distance (i.e., MSE) between the model's prediction and the real beamvector. Therefore $\sigma$ is calculated as

\begin{equation}
    \sigma^* = \underset{|\sigma|_p \leq \epsilon}{\arg max}\,\,\ell(\omega,\mathbf{x}+\sigma,\mathbf{y})
\end{equation}
where $\mathbf{y} \in \mathbb{R}^m$ is the label (i.e., beamforming vectors), and $p$ is the norm value and it can be $0, 1, 2, \infty$.

Methods for constructing adversarial examples can be categorized into groups: gradient-based, and content-based attacks, respectively \cite{vardhan2021ensemble}. In this study, gradient-based attacks were chosen as adversarial attacks because of their simplicity and variety. These attacks use the gradient of the loss
function to generate adversarial examples, i.e., incorrectly labeled. These adversarial attack types are given as follows. 
\begin{enumerate}
\def\labelenumi{(\roman{enumi})}
\item
Fast Gradient Sign Method (FGSM): FGSM is a simple, fast, and single-step attack type, which can quickly generate adversarial examples. It was first introduced by Goodfellow et al. in 2014 \cite{michels2019vulnerability}. The
gradient sign is computed using backpropagation and is quite fast. In this method, the noise, i.e., different from random noise, is added to data in the same direction (+/-) along with the loss function. The noise is adjusted by epsilon  $(\epsilon)$, which is a small number controlling the size of an adversarial attack. We can summarize the FGSM using the following equation:
\begin{equation}
	\mathbf{x}^{adv} = \mathbf{x} + \epsilon \cdot sign(\nabla_x \ell(\mathbf{\omega},\mathbf{x},y))
\end{equation}
\item
Basic Iterative Method (BIM):  BIM is an extension of the FGSM single-step attack. In this method, adversarial examples are updated by iterating many times. However, this increases the computing cost and the complexity. Unlike FGSM, BIM
manipulates the selected input with a smaller step size iteratively, and each value is calculated as in $(\epsilon)$ the neighborhood of the original input \cite{9259112}. It takes an iterative approach by applying FGSM multiple times to a small step size  \(\alpha\) instead of taking one large step, i.e.,  \(\epsilon\)/\(\alpha\). We can summarize the BIM using the following equations:
\begin{equation}
\begin{split}
    \mathbf{x}^{adv}_{0} & = \mathbf{x}, \\
    \mathbf{x}^{adv}_{N+1} & = Clip_{\mathbf{x},\epsilon} \{ \mathbf{x}^{adv}_{N} + \epsilon \cdot sign(\nabla_x \ell(\mathbf{\omega},\mathbf{x}^{adv}_{N},y)) \}
\end{split}
\end{equation}
\item
Projected Gradient Descent (PGD): The PGD attacks are similar to FGSM and BIM attack types. However, it has a different method to generate adversarial examples. It initials the search for the adversarial example at random points in a suitable region, then runs several iterations to find an adversarial example with the greatest loss, but the size of the perturbation is smaller than a specified amount referred to as epsilon, \(\epsilon\) \cite{jiang2021project}. PGD can generate stronger attacks than FGSM and BIM. 
\item
Momentum Iterative Method (MIM): MIM is a variant of the BIM adversarial attack, introducing momentum term and integrating it into iterative attacks. It improves the convergence of BIM to stabilize the direction of the gradient at each step \cite{fostiropoulosrobust}. The step size of the  \(\epsilon\) also determines the attack level of MIM as an attack parameter.
\end{enumerate}

\begin{figure*}[htbp!]
    \centering
    \includegraphics[width=1.0\linewidth]{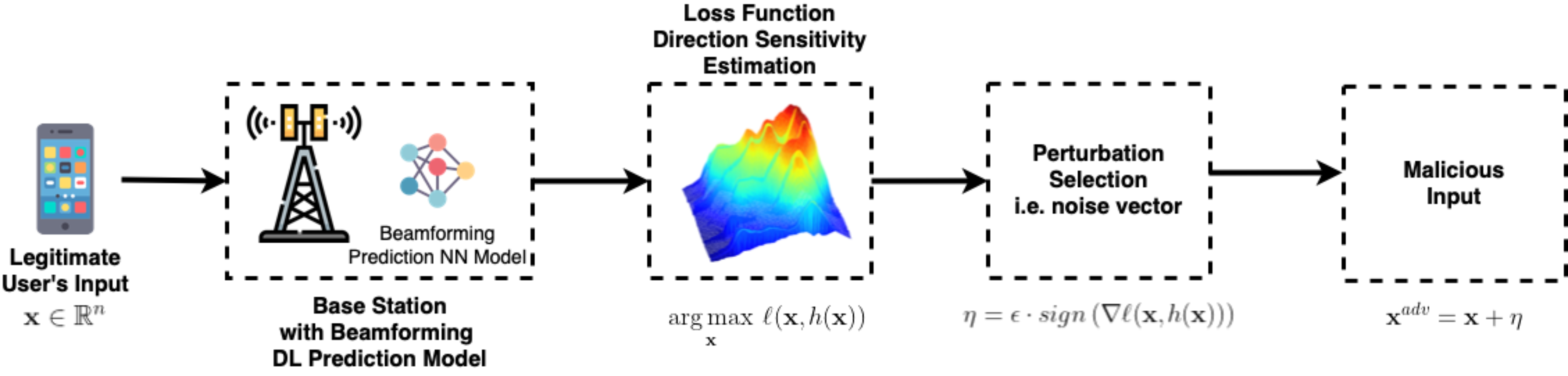}
    \caption{Typical adversarial machine learning-based malicious input generation}
    \label{fig:typical_attack}
\end{figure*}

Figure \ref{fig:typical_attack} shows a typical adversarial machine learning-based malicious input generation process.

\subsection{Mitigation Methods for wireless networks}
The DL-based beamforming prediction is vulnerable to adversarial machine learning attacks in wireless networks. Adversarial training and defensive distillation are two existing mitigation methods for adversarial machine learning attacks that also mitigate wireless communication networks.

\subsubsection{Adversarial Training}
The first mitigation method is iterative adversarial training. In this approach, the classifier is trained with the regular training data, and then the classifier is trained with the adversarial examples using the correct labels. The classifier is trained multiple times with the regular and adversarial examples. The iterative adversarial training attempts to minimize the adversarial samples' effect on the training process. However, iterative adversarial training is not efficient in practice. To obtain a robust model, the victim model must be trained with all attack types and different parameters of these attack types. Therefore, the training period of the model can be quite long.

Algorithm \ref{alg:adv_train} shows the pseudo-code of adversarial training.

\begin{algorithm}
\caption{Iterative adversarial training based mitigation}\label{alg:adv_train}
\hspace*{\algorithmicindent} \textbf{Input} $h$: vulnerable model, $\Omega$: attacks, $\Pi$: epsilon values, $\mathbf{x}_{train}$: training data, $\mathbf{y}_{train}$ training data output , $x_{test}$: test data, $y_{test}$: test data output\\
 \hspace*{\algorithmicindent} \textbf{Output} $\hat{h}$: robust model
\begin{algorithmic}[1]
\FOR{$\epsilon \in \Pi$} 
\FOR{$attack \in \Omega$} 
   \STATE $\mathbf{x}^{adv} \gets attack(\mathbf{x}_{train}, \epsilon)$ 
   \STATE $\mathbf{x}^{adv\_train} \gets \mathbf{x} \bigcup \mathbf{x}^{adv}$ 
   \STATE $h.fit(\mathbf{x}^{adv\_train}, \mathbf{y}_{train})$
\ENDFOR
\ENDFOR
\end{algorithmic}
\end{algorithm}

\subsubsection{Defensive Distillation}
\begin{figure*}[!htbp]
 \centering
	\includegraphics[width=0.95\linewidth]{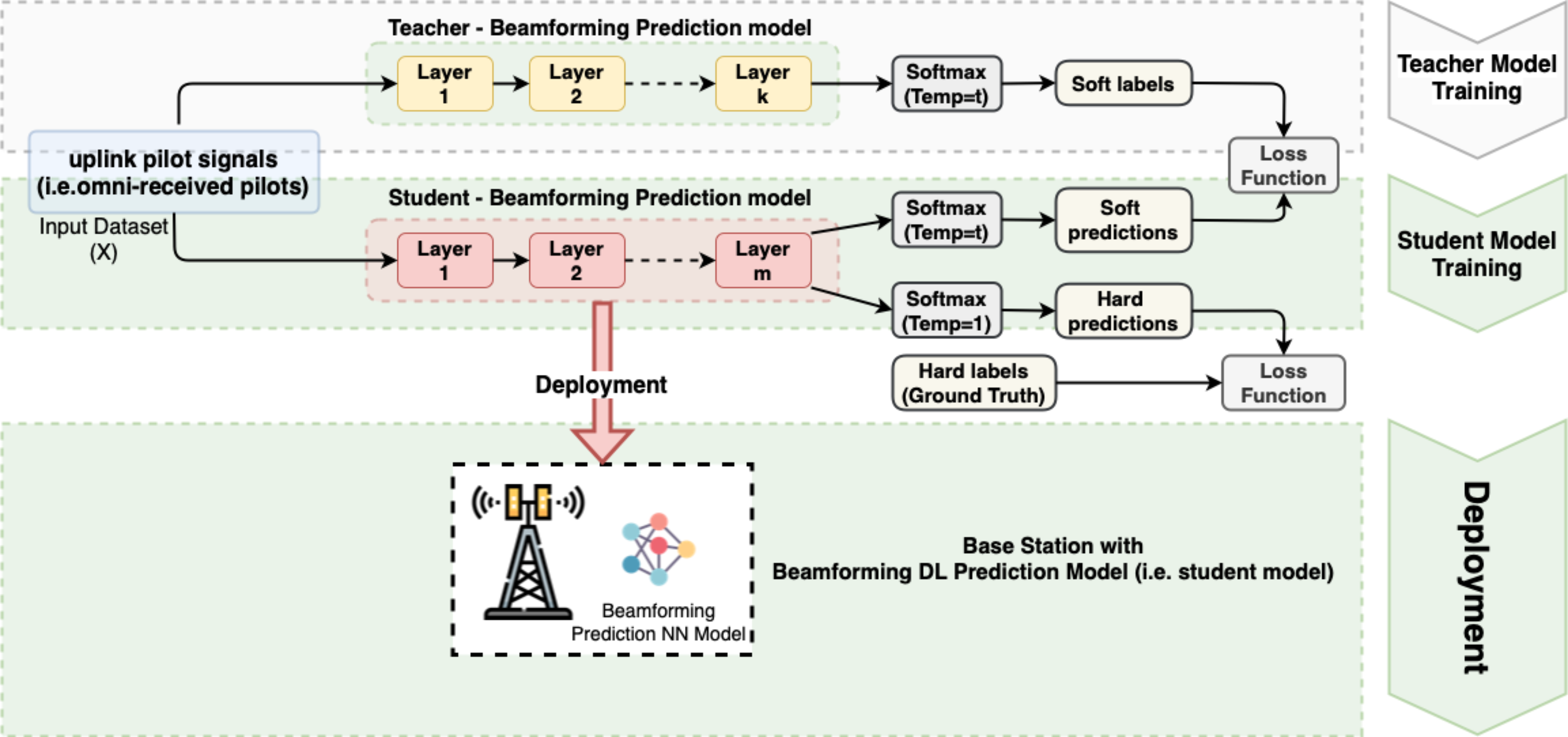}
	\caption{Defensive Distillation.}
	\label{fig:edefens-dist}
\end{figure*}

Knowledge distillation was previously introduced by Hinton et al. \cite{hinton2015distilling} to compress a large model into a smaller one. Papernot et al. \cite{papernot2016distillation} proposed this technique for the adversarial machine learning defense against attacks. The defensive distillation mitigation method includes a larger teacher model and a compressed student model. The first step is to train the teacher model with a high temperature ($T$) parameter to soften the softmax probability outputs of the DNN model. Equation \ref{eq:softmax_T} shows the modified softmax activation function as follows:
\begin{equation}
 p_{i}  = \frac{\exp(\frac{z_{i}}{T})}{\sum_{j} \exp(\frac{z_{i}}{T})} 
 \label{eq:softmax_T}
\end{equation}where $p_{i}$ is the probability of $i$\textsuperscript{th} class and $z_{i}$ are the logits.
The second step is to use the previously trained teacher model to obtain the soft labels of the training data. In this step, the teacher model predicts each of the samples in the training data using the same temperature ($T$) value, and the predictions are the labels (i.e., soft labels) for the training data to train the student model. The student model is trained with the soft labels acquired from the teacher model, again with a high $T$ value in the softmax. After the student model's training phase, the $T$ parameter is set to 1 during the prediction time of the student model. Figure \ref{fig:edefens-dist} shows the overall steps for this technique.

In the figure, the training of the beamforming prediction model (i.e., student model) is protected. against adversarial machine learning attacks and its use in base stations are shown in all its stages. The teacher model is trained as the first step, the student model is trained with the predictions made by the teacher model, and the real labels with the student model's predictions are used as the loss function inputs as the second step. In this way, the knowledge of the teacher model is compressed and transferred to the student model. The student model is deployed to the base stations in the last stage.

This technique significantly reduces the effects of gradient-based untargeted attacks. Because defense distillation has the effect of lowering the gradients down to zero, the usage of the standard objective function is no longer practical. 

\subsection{Dataset Description and scenarios} \label{sec:preliminaries}
The generic DL dataset generation framework for millimeter-wave and massive MIMO
channels (DeepMIMO) is used in experiments \cite{alkhateeb2019deepmimo}. This framework consists of two parts: (i) creating the DeepMIMO channels based on accurate ray-tracing data obtained from Wireless InSite simulator, developed by Remcom \cite{RemCom} for mmWave and massive MIMO models, and (ii) configuring a generic (parametrized) system and channel parameters to generate DeepMIMO dataset for the different applications. The ray-tracing simulation is used to generate channels based on the geometry-based characteristics. They include primarily (1) the correlation between the channels at different locations, and (2) the dependence on the environment geometry/materials. The generic (parametrized) dataset allows researchers to tune several parameters, such as the number of BSs, users, antennas, and channel paths, the system bandwidth, number of subcarriers, etc.

In this study, the DeepMIMO dataset is described for three ray-tracing scenarios, i.e., O1\_60 (outdoor - 60 GHz), I1\_2p5 (indoor - 2.5 GHz), and I3\_60 (indoor - 60 GHz). A short description of each original scenario is given as follows:
\begin{enumerate}
\def\labelenumi{(\roman{enumi})}
\item
O1\_60 is an outdoor scenario of two streets and one intersection, which includes  18 base stations supporting more than a million users. Its operating frequency is 60 GHz  \cite{O1_s}. 
\item I1\_2p5 is an indoor distributed massive MIMO scenario of a 10x10x5 (m) room with two conference tables, which includes  64 distributed antennas in the ceiling, at 2.5 m height. It can support more than 150 thousand users and, its operating frequency is 2.5 GHz  \cite{I1_s}. 
\item
I3\_60 is an indoor conference room scenario, i.e., 10x11x3 (m) conference room with its hallways, which includes  two access points inside the conference room at 2 m height. It can support more than 118 thousand users, and its operating frequency is 60 GHz  \cite{I3_s}. 
\end{enumerate}

These scenarios are revised in terms of number of BSs/APs, and active users. The revised DeepMIMO dataset parameters are given for each scenario in Table  \ref{tab:dataset}.

\begin{table}[!htbp]
\centering
\caption{DeepMIMO dataset parameters for each scenario.}
\label{tab:datasetParameter}
\begin{tabular}{|l|c|c|c|}
\hline
 & \textbf{O1\_60 } & \textbf{I1\_2p5} & \textbf{I3\_60} \\
 \hline \hline
Number of active *BSs/APs & 4 & 1 & 2 \\ \hline
Number of active users &  54300 & 12060 & 7260 \\ \hline
Number of BS antennas & 256 & 64 & 32 \\  \hline
System bandwidth & 0.5 GHz & 0.02GHz & 0.5 GHz \\ \hline
Number of subcarriers & 1024 & 64 & 32 \\  \hline
\end{tabular}
\label{tab:dataset}
\end{table}
*BS: Base Station, AP: Access Point\\

\section{System Overview}
\label{sec:overview}

\subsection{Complex Numbers and Wireless Communication}

The number system we use in our daily life is based on a real number system. There is a well-known mathematical method called the complex number system. This is based on the properties of the real number system. The complex numbers are defined as follows:  The complex numbers are the ordered pair of real numbers, written as $a + bi$. It’s a way to represent the real numbers on a plane. Wireless communication methods are based on complex numbers. The main difference between the real number system and the complex number system is that the complex number system has more than two dimensions. The complex number system is used in digital wireless communication, especially in the modulation and demodulation of wireless signals. However, adversarial machine learning attacks try to penetrate the decision boundaries of the victim DL models using real numbers, and the final malicious inputs are in the real number domain. The complex numbers are broken into their corresponding real and imaginary parts to overcome this problem. Table \ref{tab:conversion} shows the example dataset.

\begin{table}[!htbp]
    \centering
        \caption{Example training datasets. The table on the top shows the original dataset in complex numbers. The table below shows the training dataset in real numbers.}
    \begin{tabular}{|c|c|c|c|}
\hline
 \textbf{F1} & \textbf{F2} & \textbf{F3} & \textbf{F4}\\
\hline \hline
 0.04+0.79j &  0.15+0.79j &  0.21+0.79j &  0.30+0.77j \\
-0.28-0.73j & -0.35-0.72j & -0.44-0.68j & -0.50-0.61j \\
-0.15-0.78j & -0.27-0.76j & -0.33-0.73j & -0.39-0.69j \\
-0.45+0.67j & -0.34+0.71j & -0.26+0.78j & -0.17+0.78j \\
$\vdots$ & $\vdots$ & $\vdots$ & $\vdots$ \\
-0.75-0.32j & -0.77-0.24j & -0.76-0.12j & -0.78-0.02j \\
\hline
\end{tabular}
\vspace{10pt}
{\Large $\Downarrow$}
\vspace{10pt}

    \begin{tabular}{|c|c||c|c||c|c||c|c|}
\hline
 \textbf{F1-1} & \textbf{F1-2} & \textbf{F2-1} & \textbf{F2-2} & \textbf{F3-1} & \textbf{F3-2} & \textbf{F4-1} & \textbf{F4-2}\\
\hline \hline
 0.04 & 0.79 & 0.15 & 0.79  &  0.21 & 0.79 &  0.30 & 0.77 \\
-0.28 & 0.73 & -0.35 & 0.72 & -0.44 & 0.68 & -0.50 & 0.61 \\
-0.15 & 0.78 & -0.27 & 0.76 & -0.33 & 0.73 & -0.39 & 0.69 \\
-0.45 & 0.67 & -0.34 & 0.71 & -0.26 & 0.78 & -0.17 & 0.78 \\
$\vdots$ & $\vdots$ & $\vdots$ & $\vdots$ & $\vdots$ & $\vdots$ & $\vdots$ & $\vdots$  \\
-0.75 & 0.32 & -0.77 & 0.24 & -0.76 & 0.12 & -0.78 & 0.02 \\
\hline
\end{tabular}
    \label{tab:conversion}
\end{table}

\subsection{System Model}
\begin{figure*}
    \centering
    \includegraphics[width=1.0\linewidth]{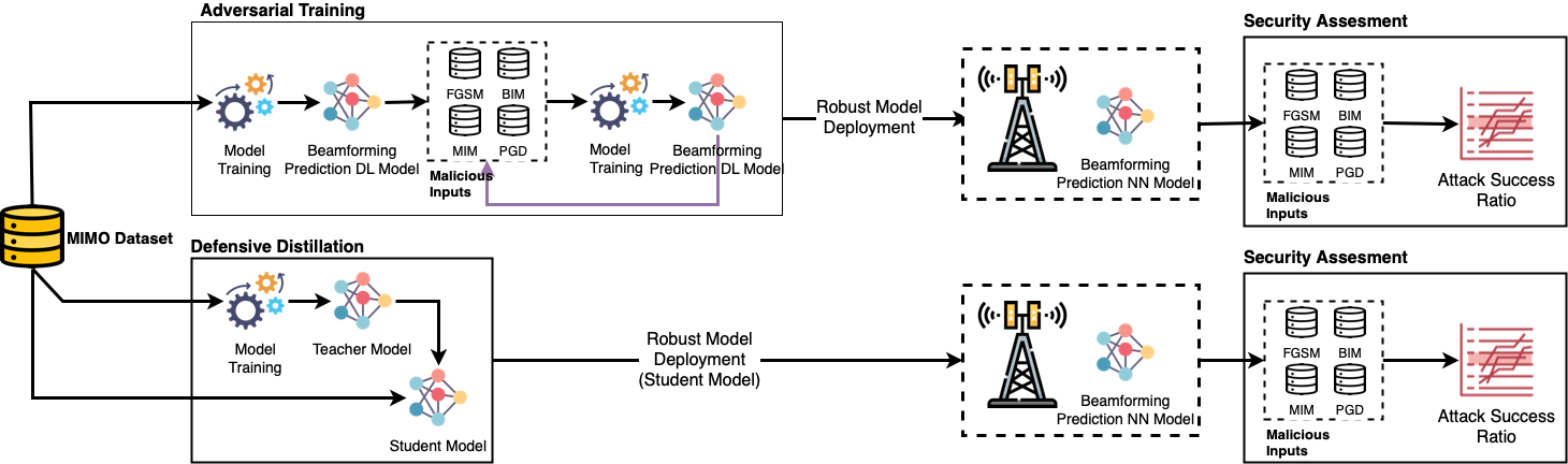}
    \caption{System overview of the proposed security scheme for beamforming prediction in 6G wireless networks.}
    \label{fig:system_flow}
\end{figure*}

In this section, a high-level system overview of the proposed security scheme for beamforming prediction in 6G wireless networks is given. The proposed security scheme is a two-phase approach: (1) adversarial training and (2) defensive distillation. In the adversarial training phase, the proposed scheme uses a modified version of the adversarial training algorithm proposed in \cite{catak2021security}. The adversarial training algorithm is used to train the deep learning models to defend against adversarial attacks. In the defensive distillation phase, the proposed scheme uses a modified version of the defensive distillation algorithm proposed in \cite{zheng2021potential}. The defensive distillation algorithm is used to improve the deep learning models against adversarial attacks. The proposed security scheme is implemented in the mmWave beamforming prediction in 6G wireless networks. Figure \ref{fig:system_flow} shows the system overview.

\section{Experiments}
\label{sec:experiments}

\subsection{Research Questions}

\begin{itemize}
    \item \textbf{RQ1}: Can we generate malicious inputs for beamforming vector prediction models using FGSM \cite{2020arXiv200702617A}, PGD, BIM, and MIM attacks in the complex domain?
    \item \textbf{RQ2}: Is there any correlation between noise vector norm value (i.e., epsilon) and prediction performance with the MSE metric?
    \item \textbf{RQ3}: What are the adversarial training and defensive distillation-based mitigation methods' protection performance metric results with different epsilon values?
\end{itemize}

\subsection{RQ1 Results}

To answer this research question, first, we train a beamforming vector prediction model on a large number of simulated data to generate realistic malicious inputs. We then apply the attack algorithms, FGSM, PGD, BIM, and MIM to generate malicious inputs and demonstrate that it is possible to generate malicious inputs for beamforming vector prediction models using the proposed attacks. Furthermore, we also examine the possibility of using the attacks for generating malicious inputs for other machine learning models. The complete paper demonstrates the feasibility of using the attacks for generating malicious inputs for beamforming vector prediction models. However, the research question of whether or not it is possible to generate malicious inputs for other machine learning models using the attacks is still open.

\begin{figure}[htbp]
    \centering
    \begin{subfigure}[b]{1.0\linewidth}
         \centering
         \includegraphics[width=1.0\linewidth]{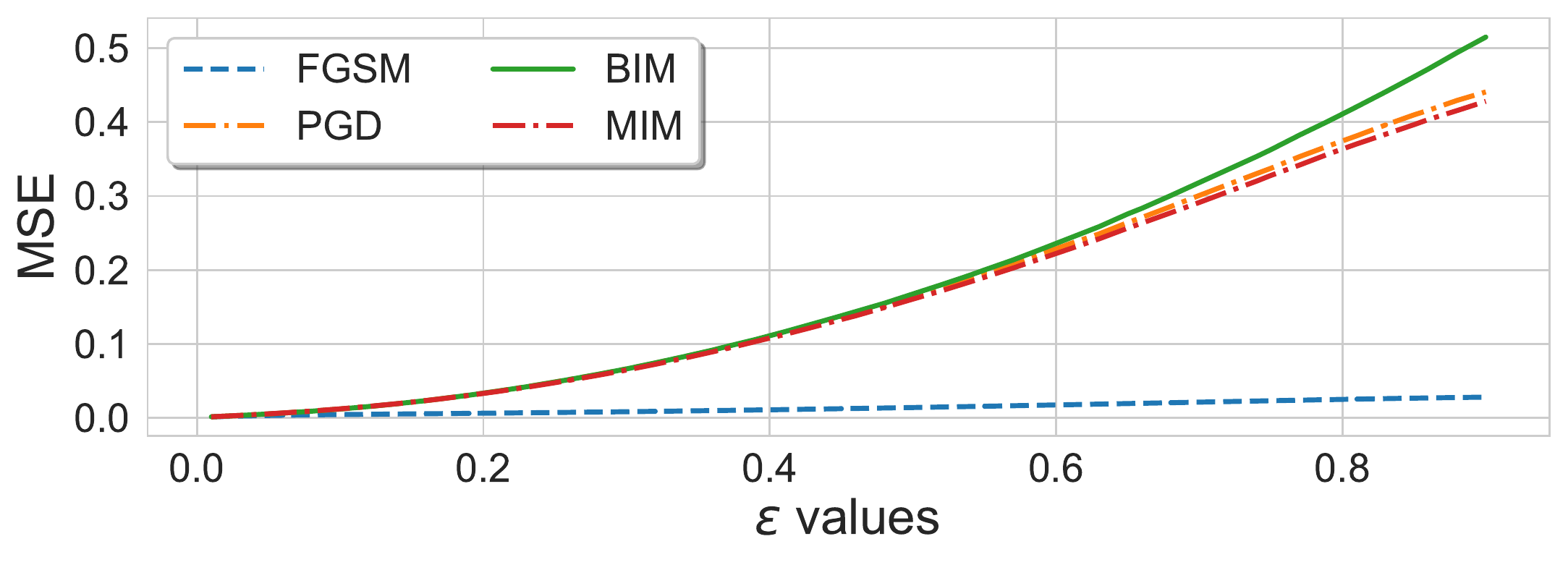}
         \caption{(a) O1\_60}
         \label{fig:O1_rq1}
    \end{subfigure}
    \begin{subfigure}[b]{1.0\linewidth}
         \centering
         \includegraphics[width=1.0\linewidth]{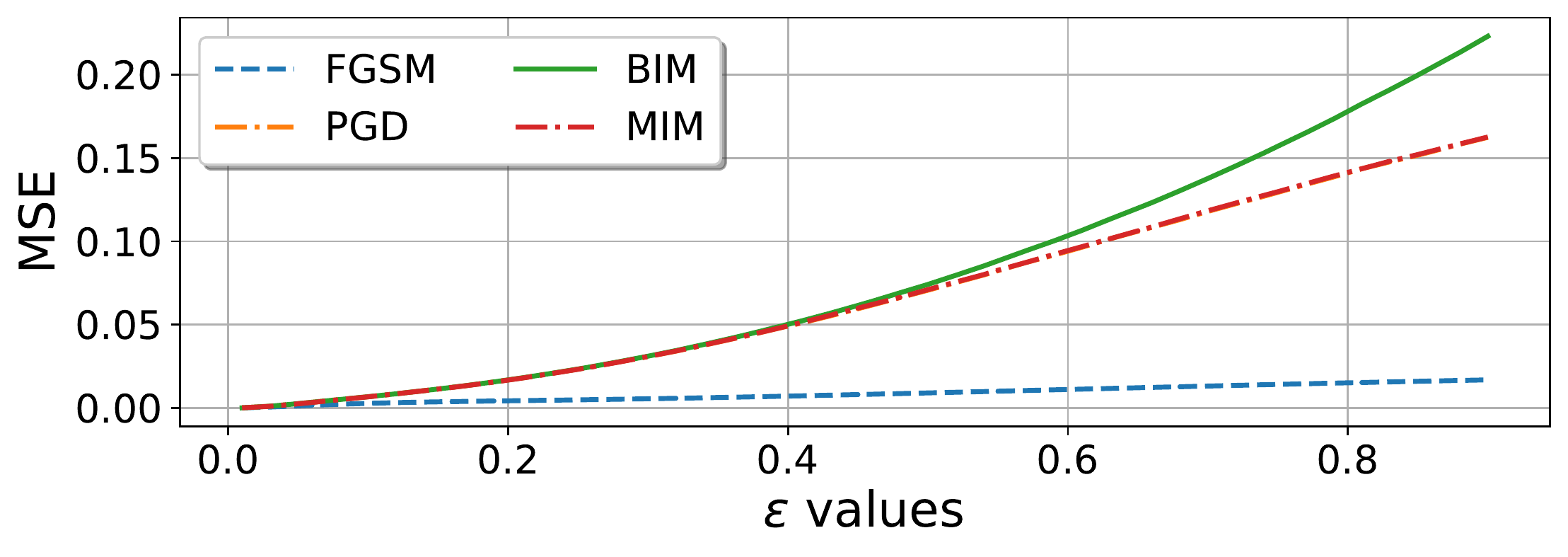}
         \caption{(b) I1\_2p5}
         \label{fig:I1_rq1}
    \end{subfigure}
    \begin{subfigure}[b]{1.0\linewidth}
         \centering
         \includegraphics[width=1.0\linewidth]{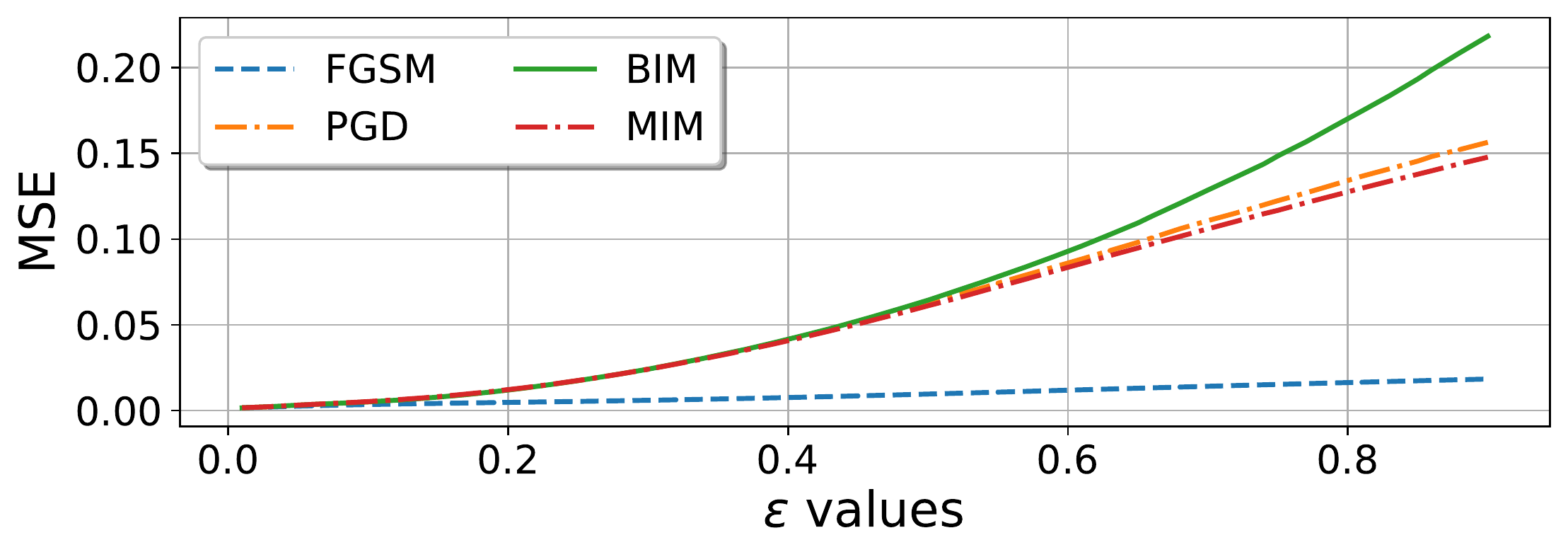}
         \caption{(c) I3\_60}
         \label{fig:I3_rq1}
    \end{subfigure}
    \caption{Prediction performance of each model using the MSE metric. The x-axis shows the $\epsilon$ budget and the y-axis shows the MSE value of the model.}
    \label{fig:rq1}
\end{figure}

Figure \ref{fig:rq1} shows the prediction performance of the beamforming vector prediction models when the malicious inputs are generated using different attack algorithms in a simulation study.

The figure shows that all the attack algorithms can generate malicious inputs for the beamforming vector prediction model. As we can see from the figure, the BIM attack has the highest prediction error rate (i.e., attack success ratio). The PGD attack has the second-highest prediction error rate. The prediction error rate of the MIM attack has the third-highest prediction error rate. The FGSM attack has the the lowest. From the figures, one can say that the beamforming vector prediction models are more sensitive to the BIM attack, whereas the FGSM attack is less sensitive to the predictions.\\

\fbox{\begin{minipage}{1.0\linewidth}
\textbf{Concluding Remarks for RQ1:} The attackers can generate malicious inputs for the beamforming vector prediction model. From the attacker's perspective, the most successful attack is BIM.
\end{minipage}}

\subsection{RQ2 Results}

We examined the correlation between noise vector norm value (i.e., epsilon) and prediction performance with MSE metric in a simulation study. The results show that the prediction performance with the MSE metric is strongly correlated with the noise vector norm value when noise is added to the input features and the target feature.

\begin{table*}[!htbp]
\centering
\caption{Pearson correlation coefficients of the scenarios and attacks.}
\label{tab:rq2}
\begin{tabular}{|l|c|c|c|c|c|c|c|c|}
\hline
& \multicolumn{2}{c|}{\textbf{FGSM}} & \multicolumn{2}{c|}{\textbf{PGD}} & \multicolumn{2}{c|}{\textbf{BIM}} & \multicolumn{2}{c|}{\textbf{MIM}} \\
\cline{2-9}
& $r$ & $p$ & $r$ & $p$ & $r$ & $p$ & $r$ & $p$\\ \hline \hline
O1\_60 & 0.99109 & 0.0 & 0.97996 & 0.0 & 0.97182 & 0.0 & 0.98024 & 0.0\\
I1\_2P5 & 0.99585 & 0.0 & 0.99178 & 0.0 & 0.97613 & 0.0 & 0.99208 & 0.0\\
I3\_60 & 0.99258 & 0.0 & 0.97448 & 0.0 & 0.95398 & 0.0 & 0.97632 & 0.0 \\ \hline
\end{tabular}
\end{table*}

This simulation study examines the correlation between noise vector norm value (i.e., epsilon) and prediction performance with the MSE metric. The results show that the prediction performance with the MSE metric is strongly correlated with the noise vector norm value (i.e., epsilon) when noise is added to the input and target features.  

Table \ref{tab:rq2} shows the Pearson correlation coefficients of the relation between epsilon budget and MSE value. Pearson correlation is a statistical measure of the linear correlation between two variables. It is a measure of the extent to which two variables vary together. A correlation of 1.0 means that the two variables vary completely together; a correlation of 0 means that the two variables vary entirely independently. In the case of Table \ref{tab:rq2}, the correlation coefficient of the relation between epsilon budgets and MSE value is around 0.99. This means that the prediction performance with the MSE metric is strongly correlated with the noise vector norm value (i.e., epsilon) when noise is added to the input and target features.

\begin{figure}[htbp]
     \centering
     \begin{subfigure}[b]{0.9\linewidth}
         \centering
         \includegraphics[width=0.7\linewidth]{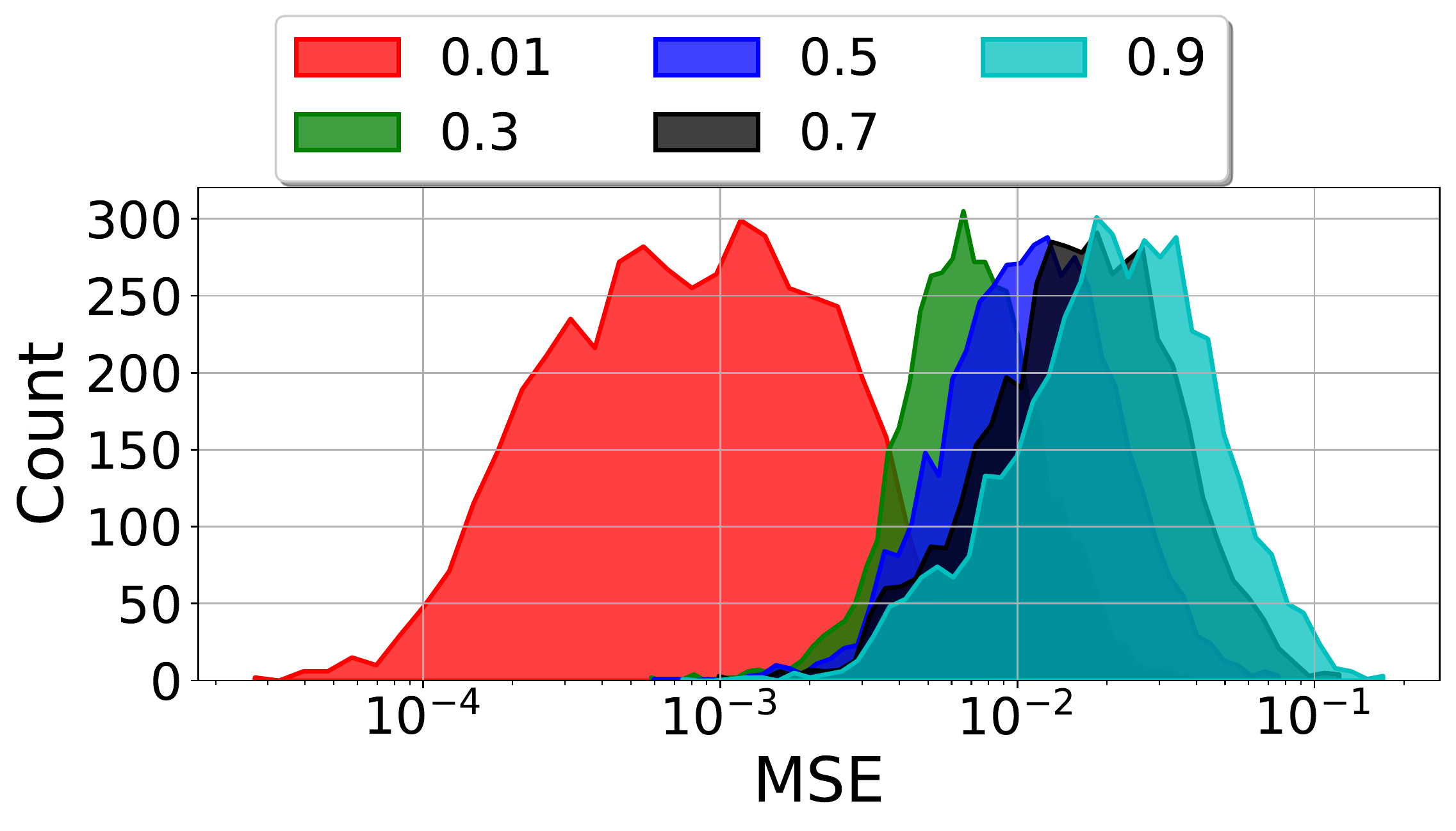}
         \caption{(a) FGSM}
         \label{fig:O1_FGSM_rq2}
     \end{subfigure}
     \hfill
     \begin{subfigure}[b]{0.9\linewidth}
         \centering
         \includegraphics[width=0.7\linewidth]{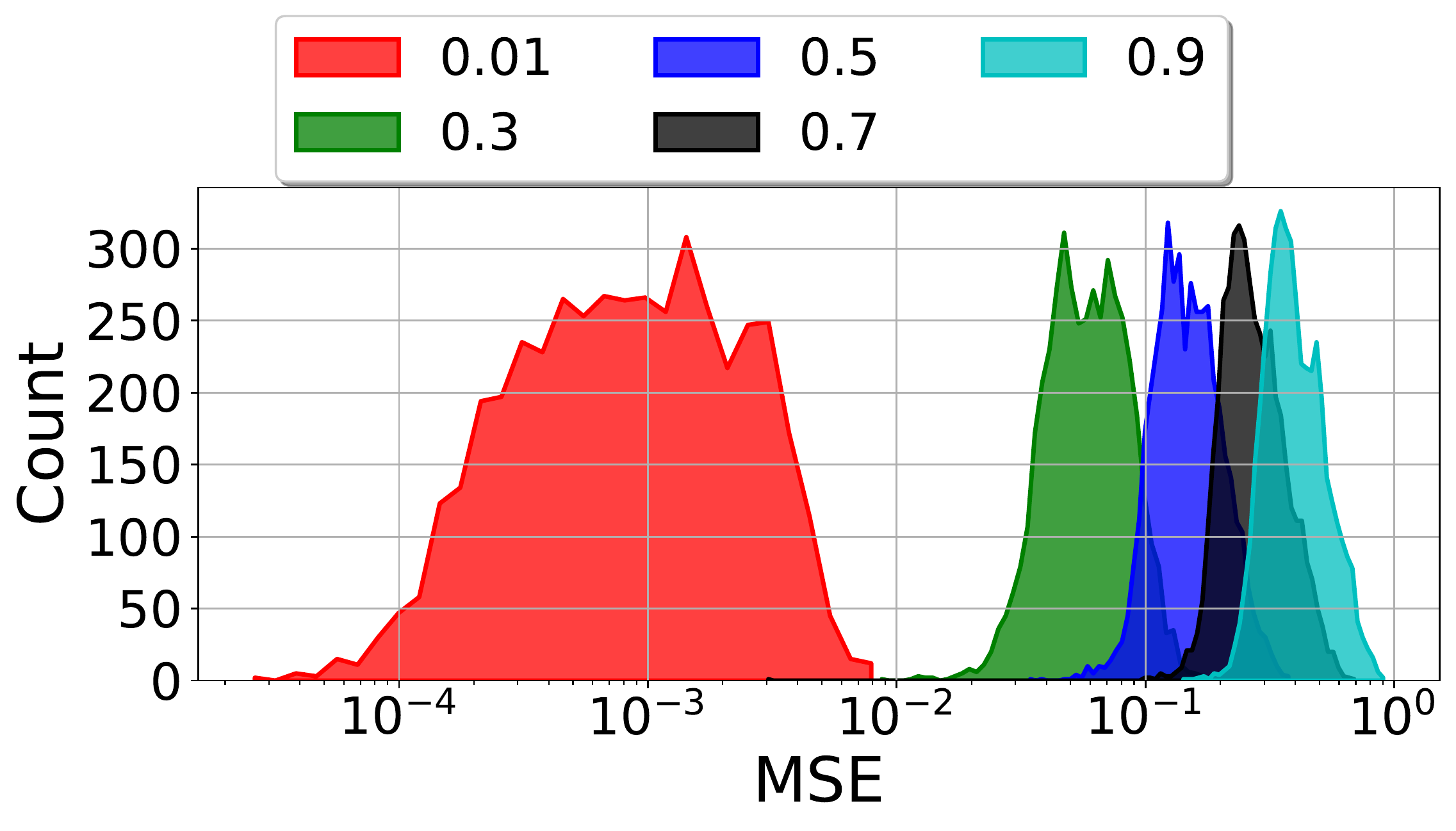}
         \caption{(b) PGD}
         \label{fig:O1_PGD_rq2}
     \end{subfigure}
     \hfill
     \begin{subfigure}[b]{0.9\linewidth}
         \centering
         \includegraphics[width=0.7\linewidth]{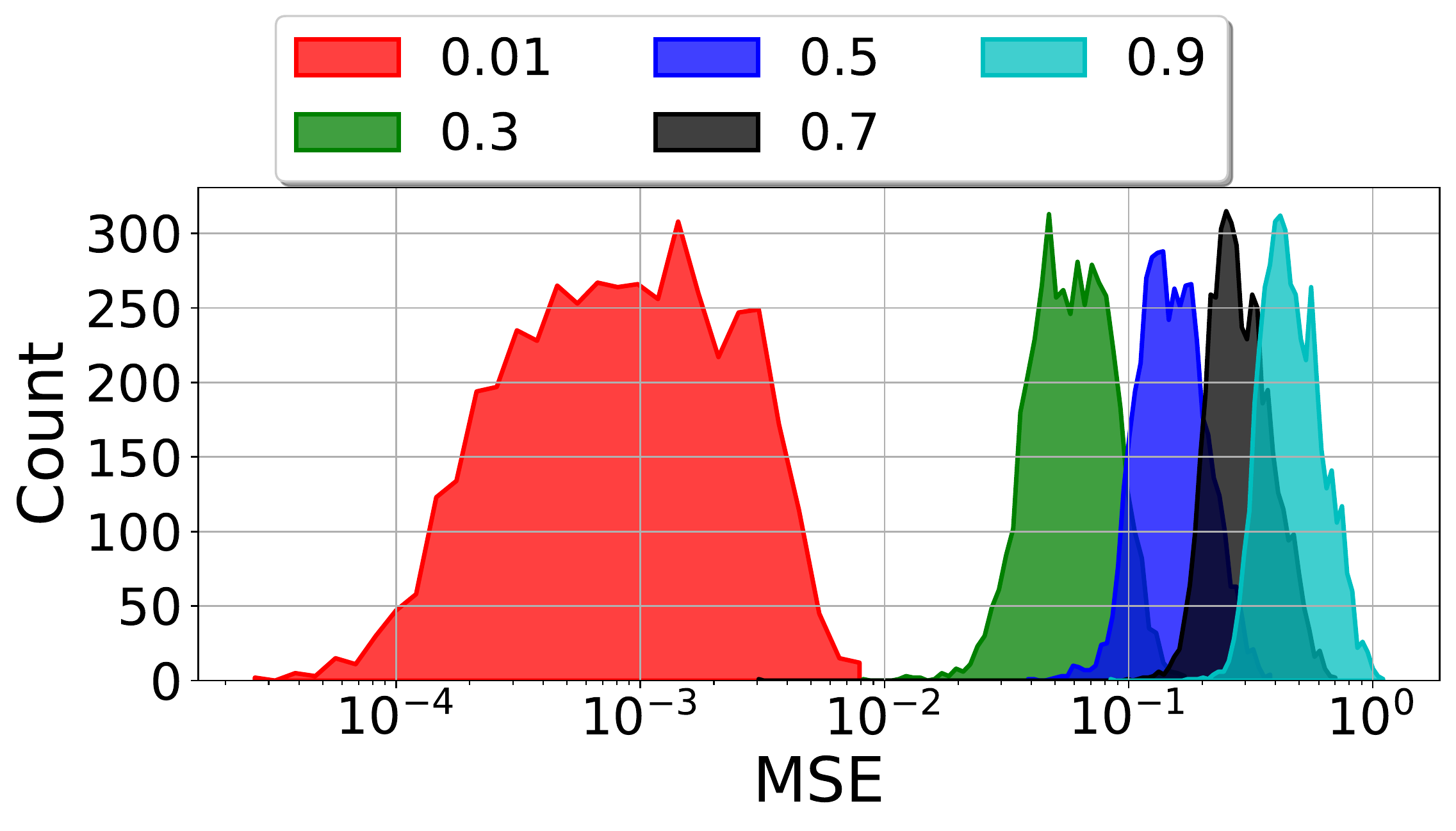}
         \caption{(c) BIM}
         \label{fig:O1_BIM_rq2}
     \end{subfigure}
     \hfill
     \begin{subfigure}[b]{0.9\linewidth}
         \centering
         \includegraphics[width=0.7\linewidth]{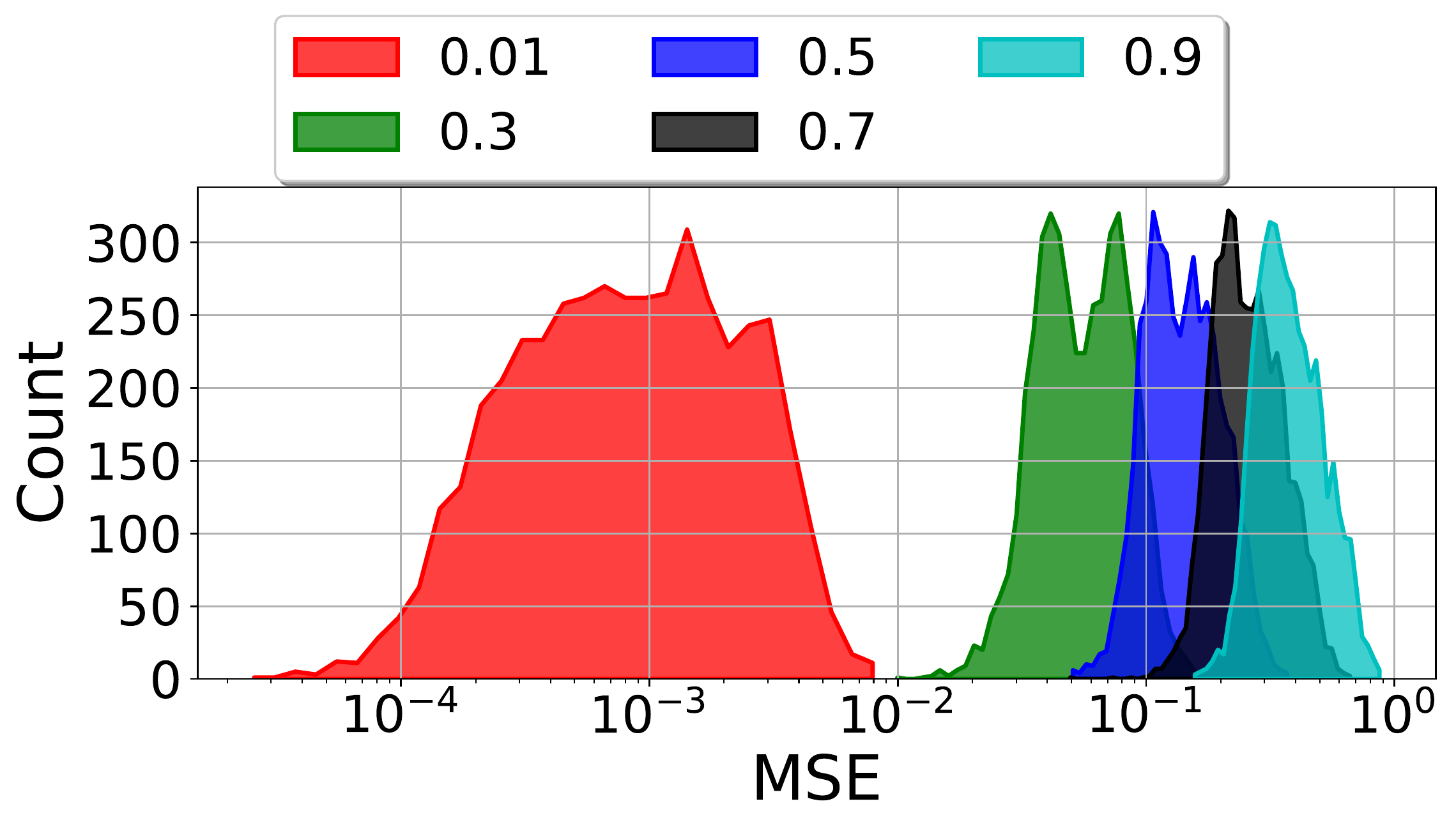}
         \caption{(d) MIM}
         \label{fig:O1_MIM_rq2}
     \end{subfigure}
        \caption{O1\_60: MSE distributions of the malicious inputs with five different epsilon values ($\epsilon \in \{0.01, 0.3, 0.5, 0.7, 0.9\}$).}
        \label{fig:O1_rq2}
\end{figure}

\begin{figure}[htbp]
     \centering
     \begin{subfigure}[b]{1.0\linewidth}
         \centering
         \includegraphics[width=0.7\linewidth]{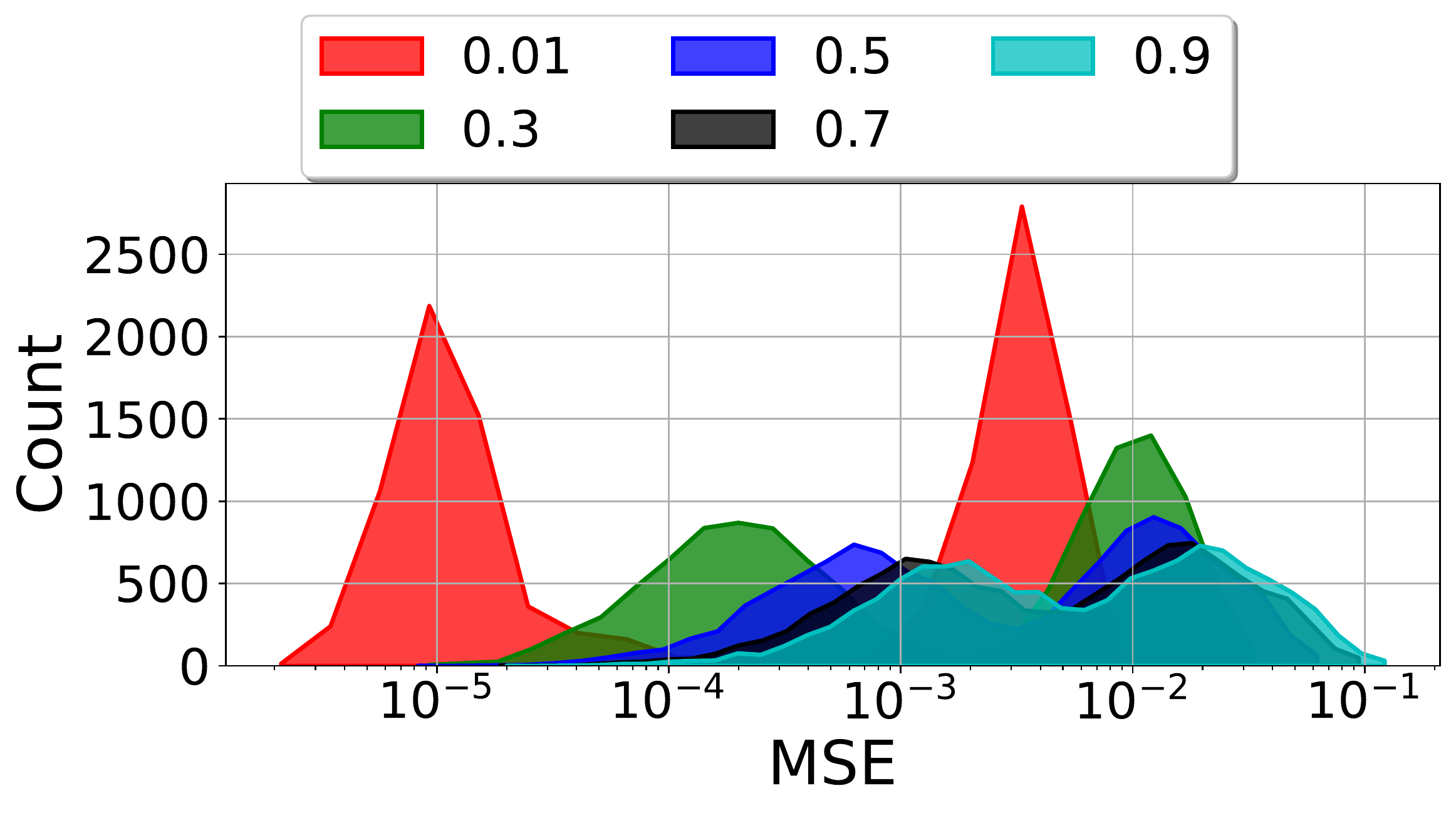}
         \subcaption{(a) FGSM}
         \label{fig:I3_60_FGSM_rq2}
     \end{subfigure}
     \hfill
     \begin{subfigure}[b]{1.0\linewidth}
         \centering
         \includegraphics[width=0.7\linewidth]{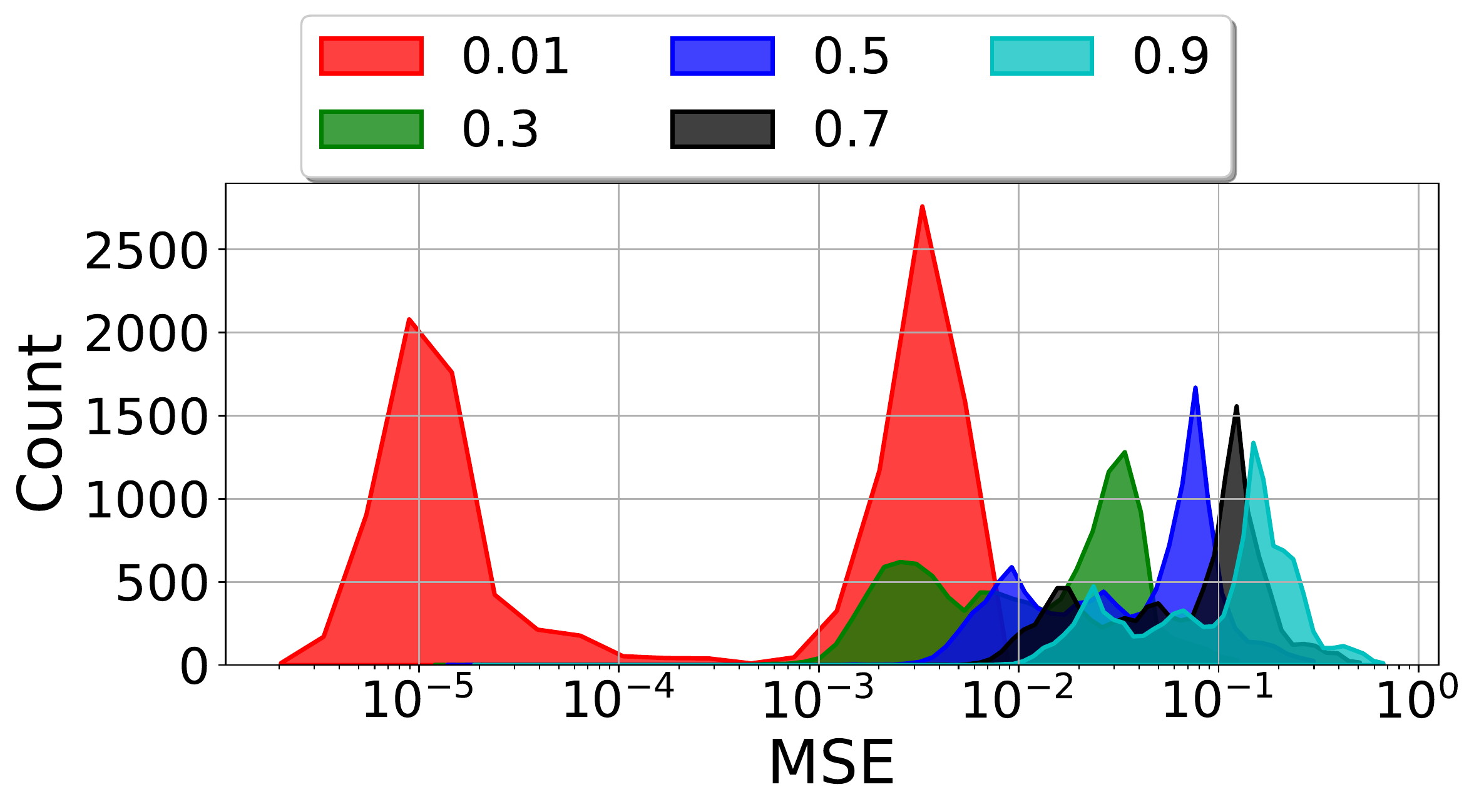}
         \subcaption{(b) PGD}
         \label{fig:I3_60_PGD_rq2}
     \end{subfigure}
     \hfill
     \begin{subfigure}[b]{1.0\linewidth}
         \centering
         \includegraphics[width=0.7\linewidth]{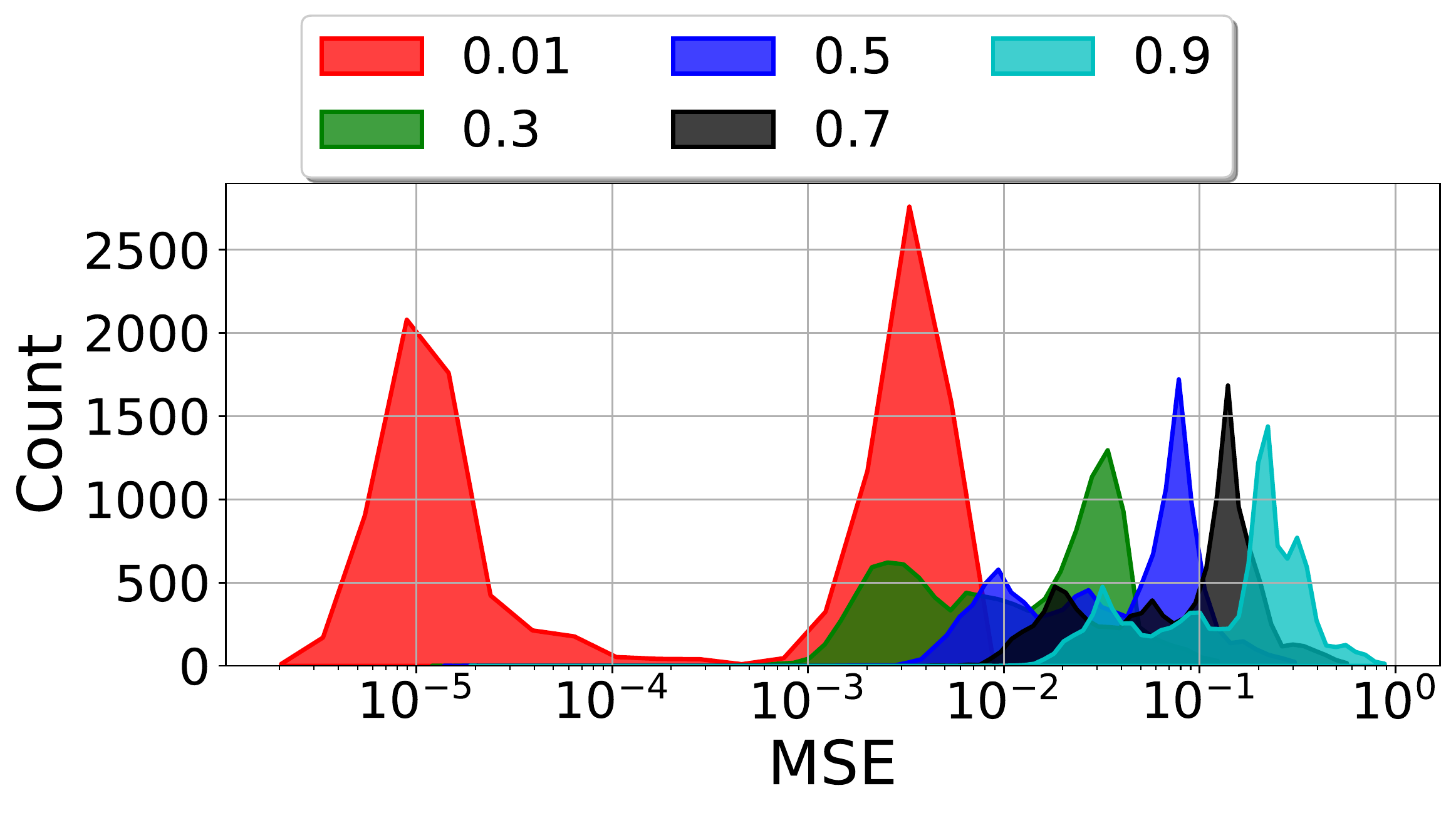}
         \subcaption{(c) BIM}
         \label{fig:I3_60_BIM_rq2}
     \end{subfigure}
     \hfill
     \begin{subfigure}[b]{1.0\linewidth}
         \centering
         \includegraphics[width=0.7\linewidth]{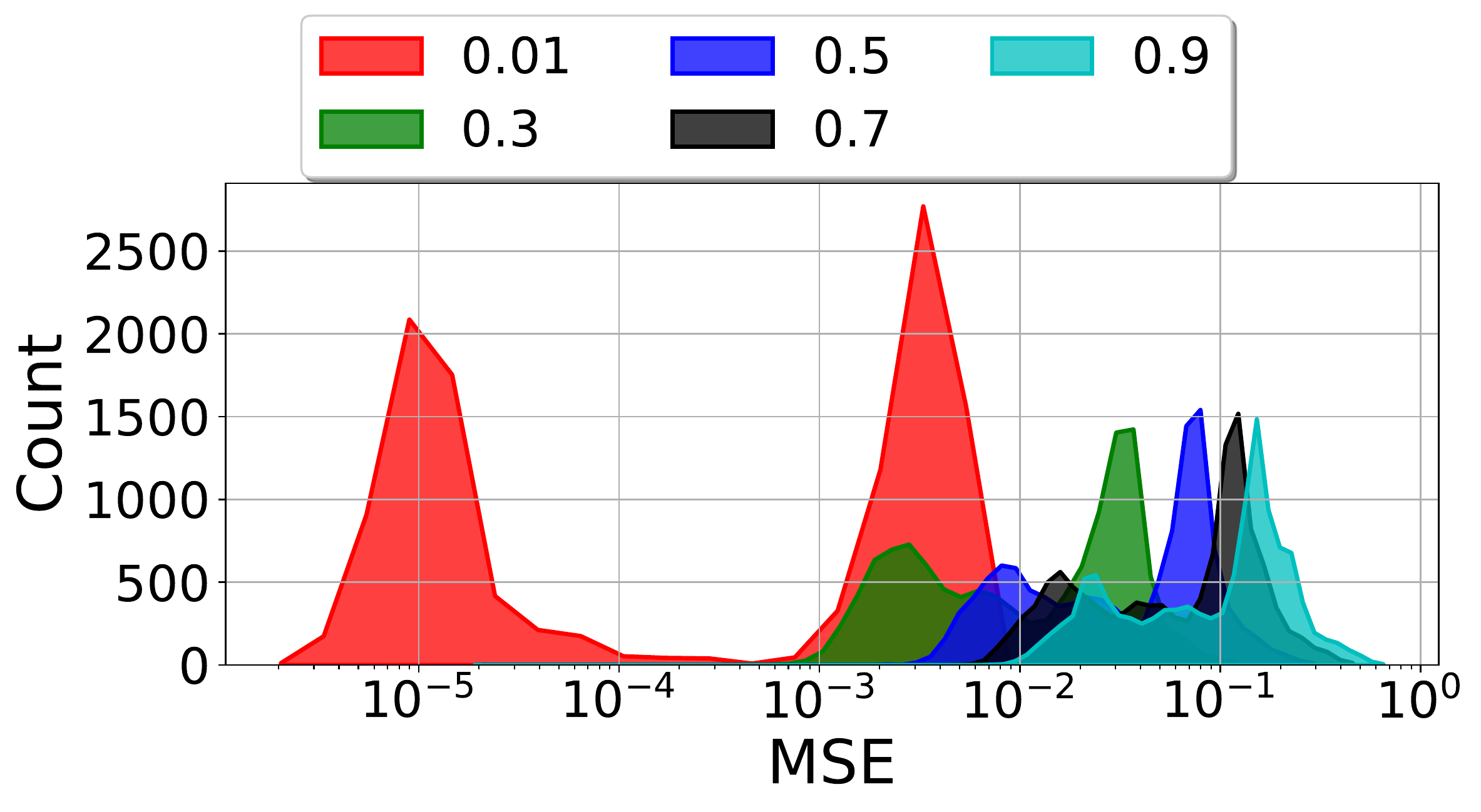}
         \subcaption{(d) MIM}
         \label{fig:I3_60_MIM_rq2}
     \end{subfigure}
        \caption{I3\_60: MSE distributions of the malicious inputs with five different epsilon values ($\epsilon \in \{0.01, 0.3, 0.5, 0.7, 0.9\}$).}
        \label{fig:I3_60_rq2}
\end{figure}

\begin{figure}[htbp]
     \centering
     \begin{subfigure}[b]{1.0\linewidth}
         \centering
         \includegraphics[width=0.7\linewidth]{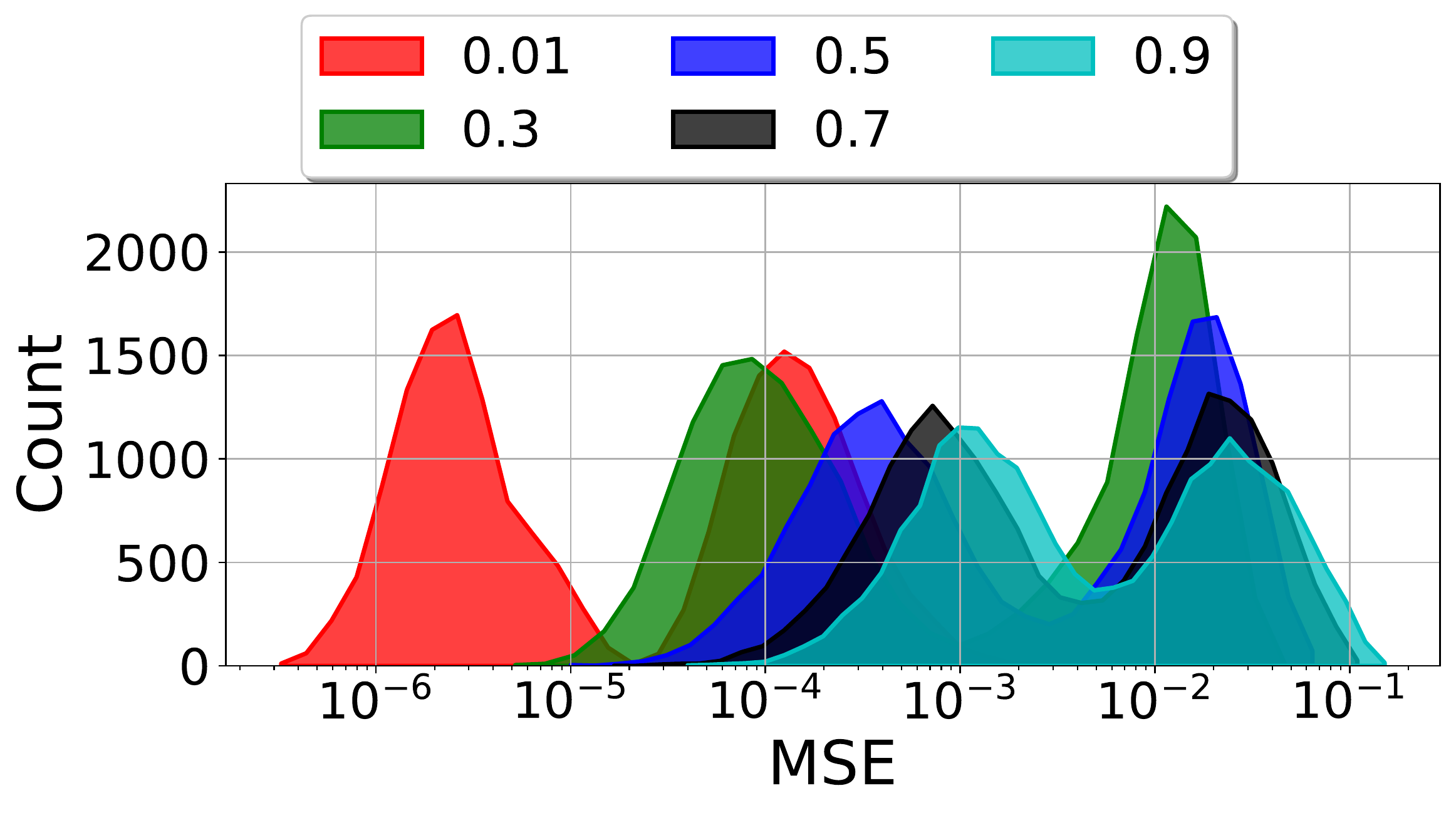}
         \subcaption{(a) FGSM}
         \label{fig:I1_2p5_FGSM_rq2}
     \end{subfigure}
     \hfill
     \begin{subfigure}[b]{1.0\linewidth}
         \centering
         \includegraphics[width=0.7\linewidth]{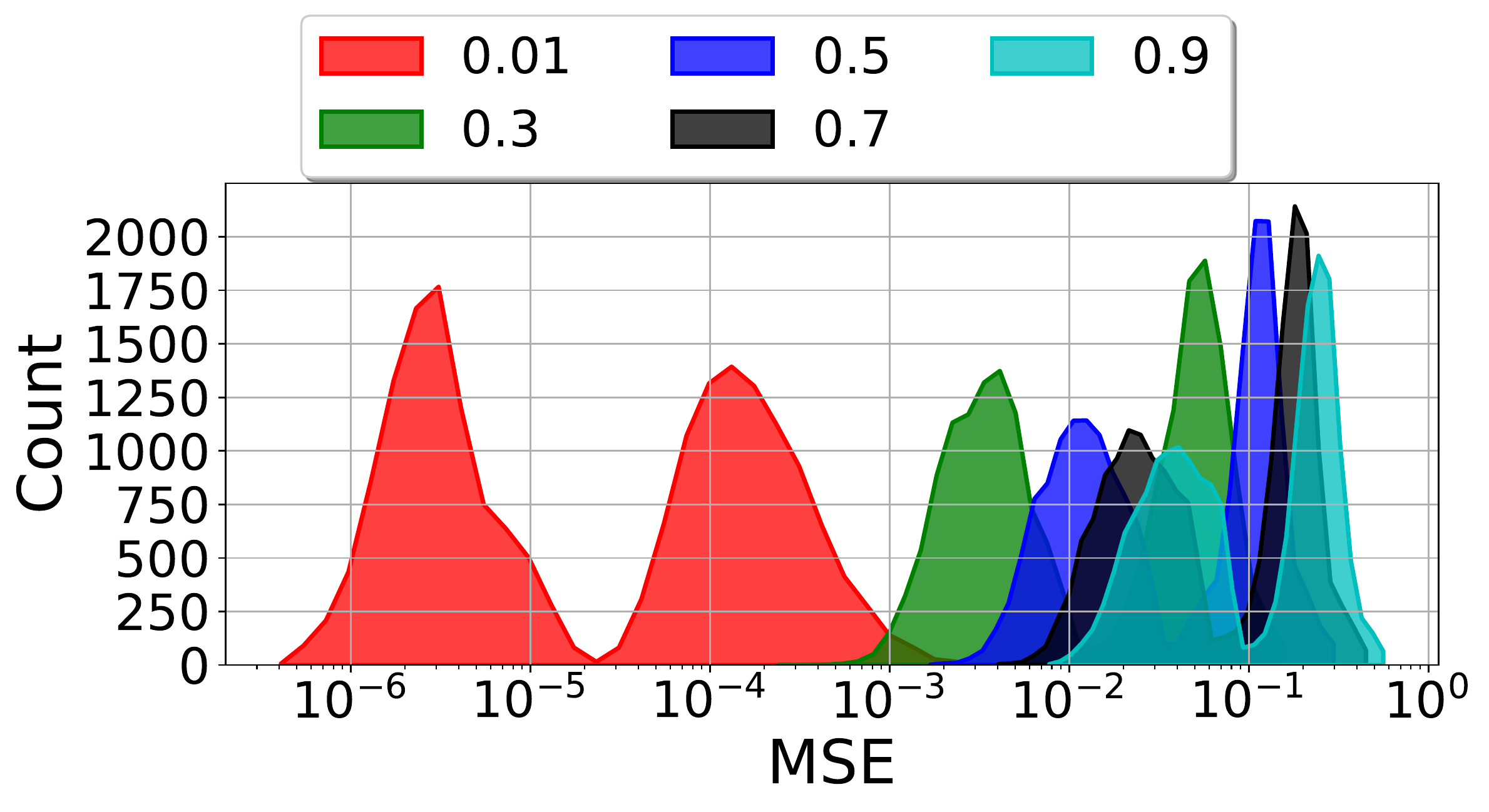}
         \subcaption{(b) PGD}
         \label{fig:I1_2p5_PGD_rq2}
     \end{subfigure}
     \hfill
     \begin{subfigure}[b]{1.0\linewidth}
         \centering
         \includegraphics[width=0.7\linewidth]{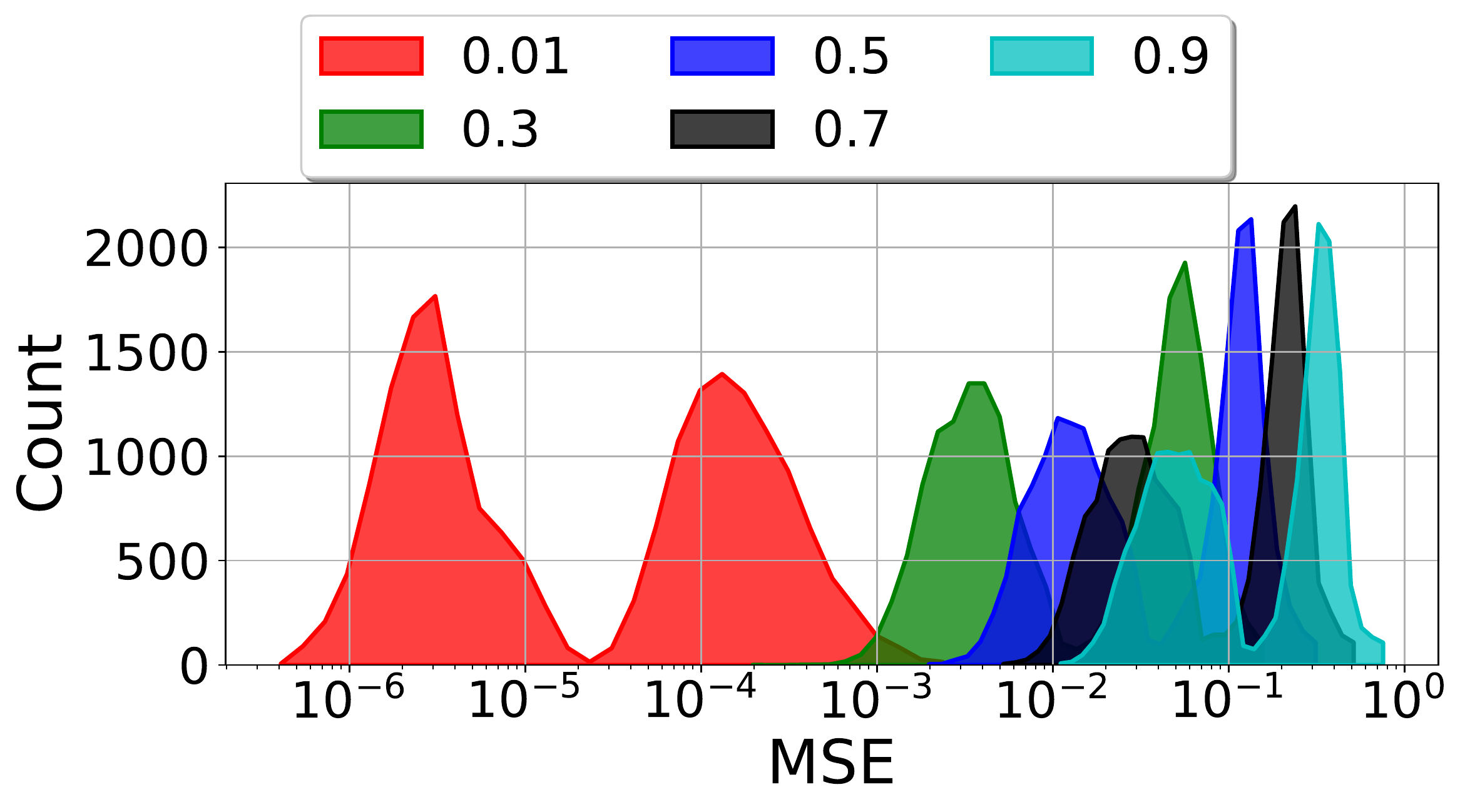}
         \subcaption{(c) BIM}
         \label{fig:I1_2p5_BIM_rq2}
     \end{subfigure}
     \hfill
     \begin{subfigure}[b]{1.0\linewidth}
         \centering
         \includegraphics[width=0.7\linewidth]{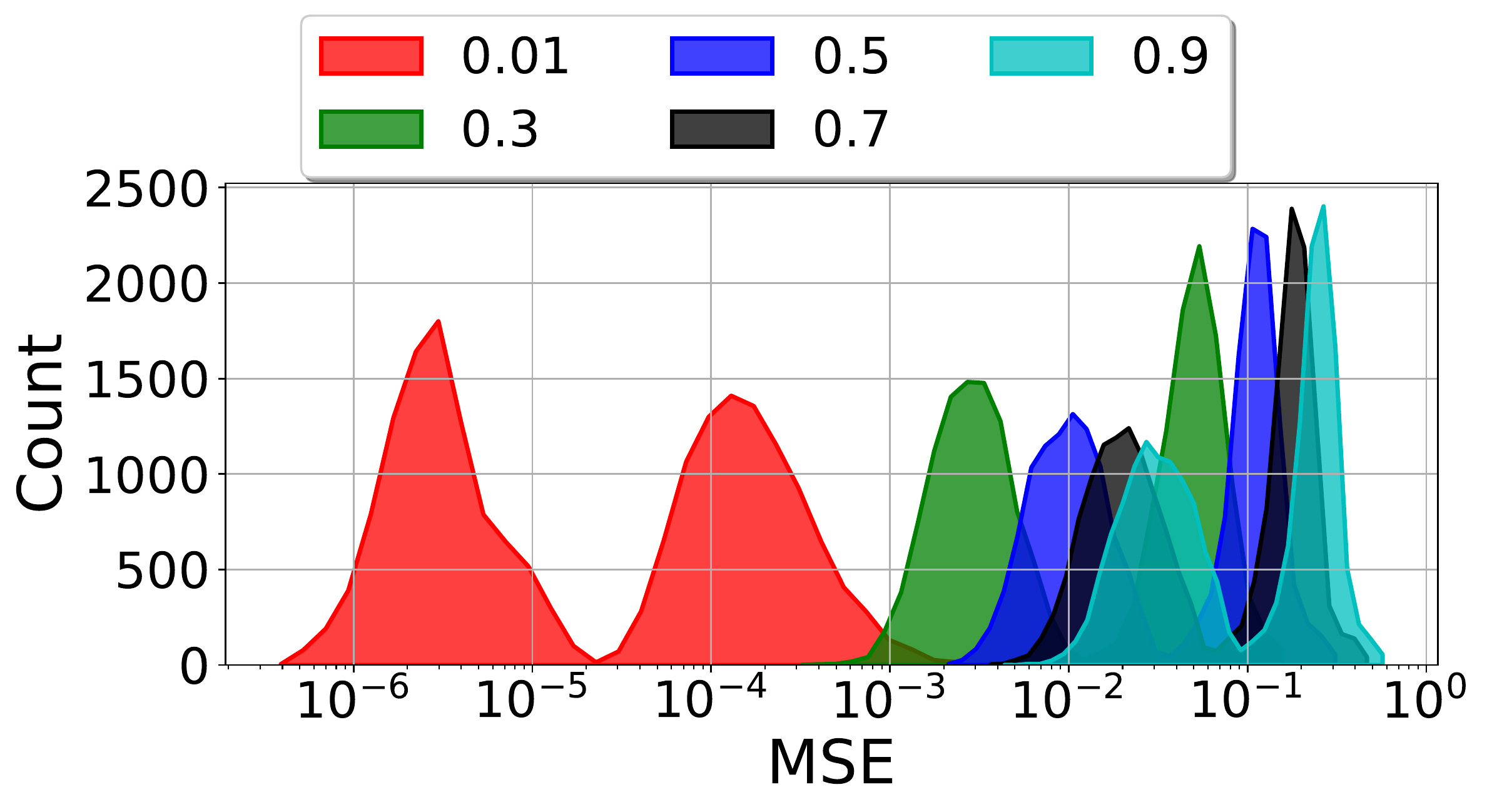}
         \subcaption{(d) MIM}
         \label{fig:I1_2p5_MIM_rq2}
     \end{subfigure}
        \caption{I1\_2p5: MSE distributions of the malicious inputs with five different epsilon values ($\epsilon \in \{0.01, 0.3, 0.5, 0.7, 0.9\}$).}
        \label{fig:I1_2p5_rq2}
\end{figure}

Figures \ref{fig:O1_rq2}-\ref{fig:I1_2p5_rq2} show the MSE distributions of each input instance in the malicious inputs generated with different attack algorithms. The figure represents the distribution of the MSE obtained from the malicious inputs generated by adding noise to the input features and the target feature. The figure indicates that the MSE distribution of the malicious inputs generated by adding noise to the input features is not uniform. In the case of the MIM attack, the figure shows that the MSE distribution of the malicious inputs generated by adding noise to the input features and the target feature has a minimal variance. This is because the MIM attack adds noise to the input and target features. The BIM attack adds noise to the input features. Thus, the MSE distribution of the malicious inputs generated by adding noise to the input features and the target feature has a more significant variance than the MIM attack. The PGD attack adds Gaussian noise to the input features. Thus the MSE distribution of the malicious inputs generated by adding noise to the input features has a more significant variance than the FGSM attack. The FGSM attack adds Gaussian noise to the input features. Thus the MSE distribution of the malicious inputs generated by adding noise to the input features has a more significant variance than the PGD attack. \\

\fbox{\begin{minipage}{1.0\linewidth}
\textbf{Concluding Remarks for RQ2:} There is a strong negative correlation between $\epsilon$ and DL model's prediction performance. The confidence interval value of the correlations (i.e., $p$-value) is 0. The $p$-value is the probability connected to the likelihood of acquiring the correlation result.
\end{minipage}}

\subsection{RQ3 Results}

\begin{figure}[htbp]
     \centering
     \begin{subfigure}[b]{1.0\linewidth}
         \centering
         \includegraphics[width=1.0\linewidth]{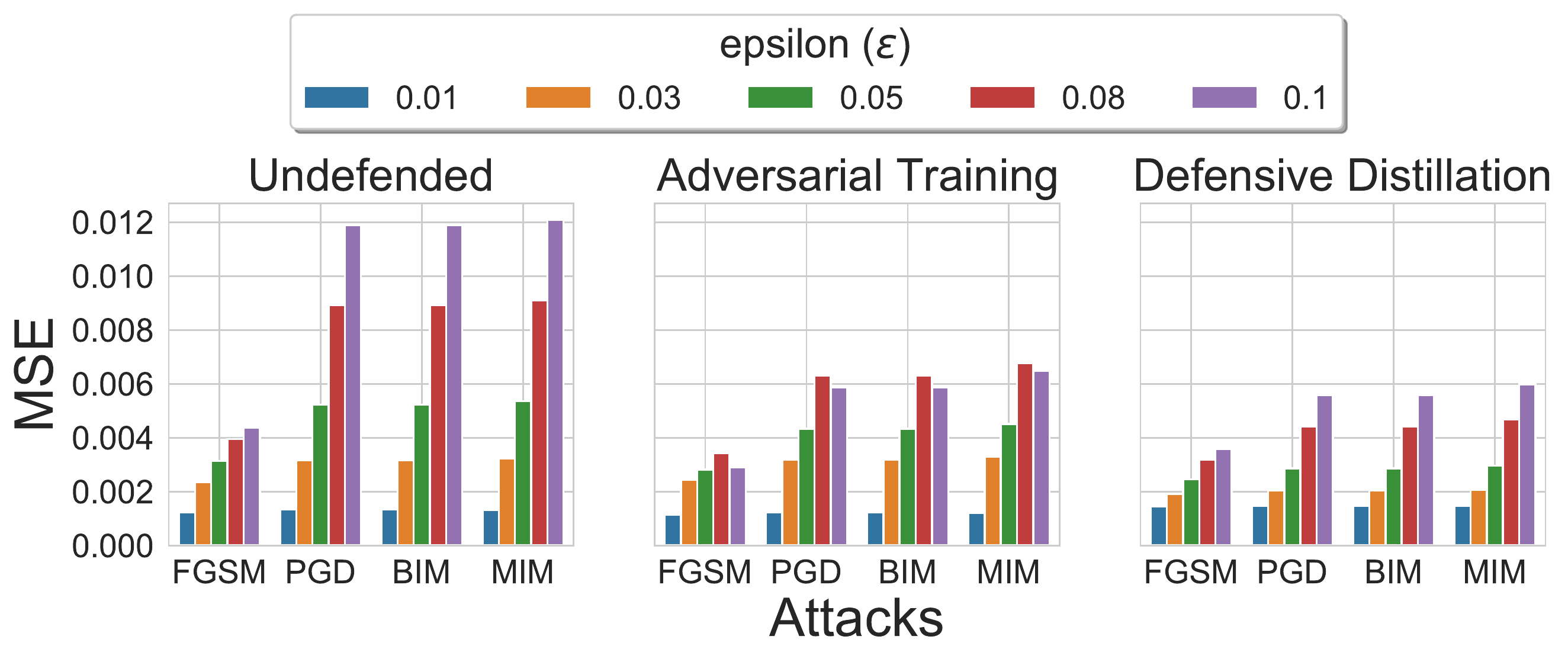}
         \subcaption{(a) O1\_60}
         \label{fig:O1_mitigations}
     \end{subfigure}
     \hfill
     \begin{subfigure}[b]{1.0\linewidth}
         \centering
         \includegraphics[width=1.0\linewidth]{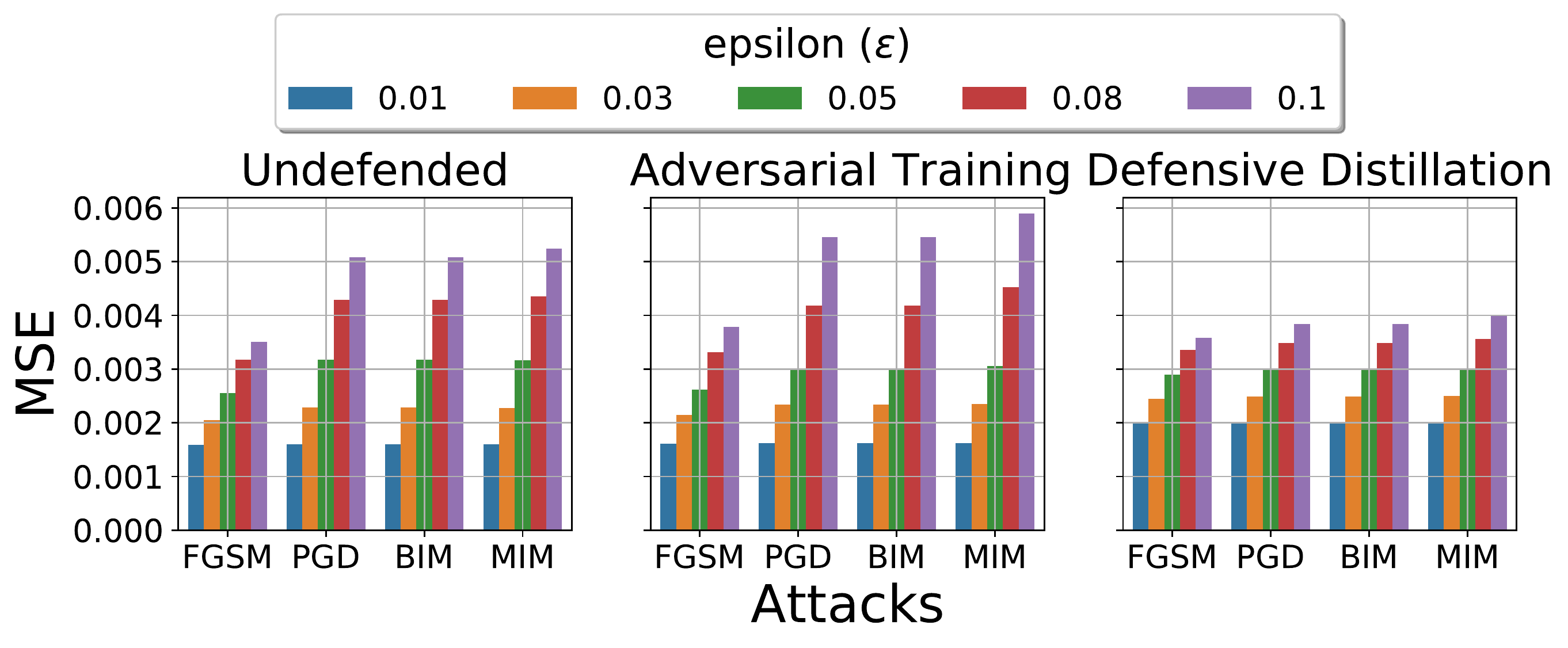}
         \subcaption{(b) I3\_60}
         \label{fig:I3_60_mitigations}
     \end{subfigure}
     \hfill
     \begin{subfigure}[b]{1.0\linewidth}
         \centering
         \includegraphics[width=1.0\linewidth]{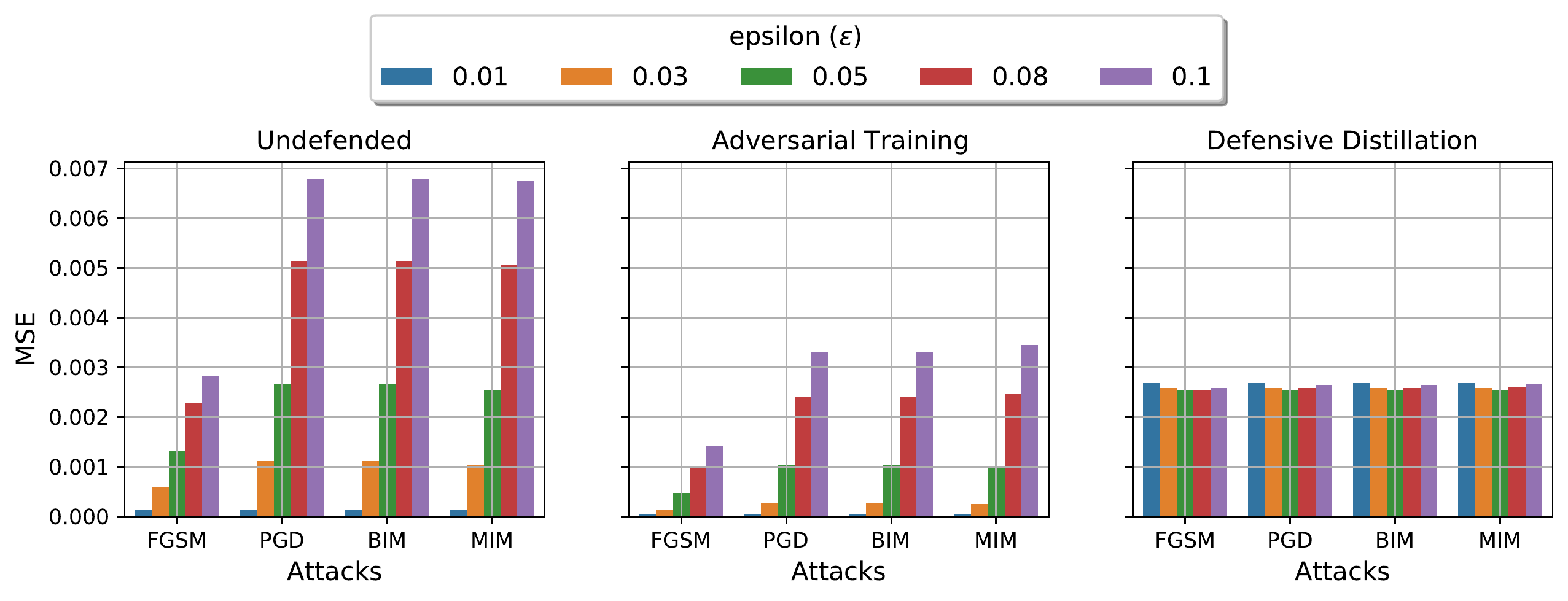}
         \subcaption{(c) I1\_2p5}
         \label{fig:I1_2p5_mitigations}
     \end{subfigure}
     \caption{Adversarial training and defensive distillation-based mitigation methods results.}
     \label{fig:rq3}
\end{figure}

Figure \ref{fig:rq3} summarizes the experiment results for the adversarial training and defensive distillation mitigation methods for different epsilon values ($\epsilon \in \{0.01, 0.03, 0.05, 0.08, 0.10\}$). Except for the I1\_2p5 scenario, the MSE values of the model, which has been made robust by the defensive distillation mitigation method, are lower (i.e., the prediction performance is higher) in the other two scenarios. In the I1\_2p5 scenario, the adversarial training method is more successful in protecting against attacks with low $\epsilon$  values (i.e., $\epsilon$ < 0.08). In contrast, in cases where the epsilon value is 0.08 or higher, the defensive distillation mitigation method creates a more successful defense.

Table \ref{tab:rq3} shows the experiment results for each scenario, attack, and mitigation method.

\begin{table}[!htbp]
    \centering
    \caption{The summary of the experiment results for each scenario, attack, and mitigation method.}
    \begin{tabular}{|l|l|rrp{1.1cm}p{1.4cm}|}
\hline
\textbf{Sc.} & \textbf{Attack} & \textbf{Epsilon} & \textbf{Undefended} &  \textbf{Adversarial Training} &  \textbf{Defensive Distillation} \\
\hline \hline
    \multirow{16}{*}{O1\_60} & \multirow{4}{*}{BIM} & 0.03 &   0.003160 &            0.003194 &       0.002033 \\
    & & 0.05 &   0.005232 &            0.004333 &       0.002864 \\
     && 0.08 &   0.008918 &            0.006315 &       0.004427 \\
     && 0.10 &   0.011888 &            0.005867 &       0.005574 \\ \cline{2-6}
   &\multirow{4}{*}{FGSM} & 0.03 &   0.002352 &            0.002440 &       0.001918 \\
    && 0.05 &   0.003137 &            0.002819 &       0.002457 \\
    && 0.08 &   0.003950 &            0.003431 &       0.003183 \\
    && 0.10 &   0.004380 &            0.002900 &       0.003591 \\ \cline{2-6}
    &\multirow{4}{*}{MIM} & 0.03 &   0.003226 &            0.003303 &       0.002062 \\
     && 0.05 &   0.005363 &            0.004510 &       0.002963 \\
     && 0.08 &   0.009107 &            0.006778 &       0.004686 \\
     && 0.10 &   0.012092 &            0.006488 &       0.005984 \\ \cline{2-6}
    & \multirow{4}{*}{PGD} & 0.03 &   0.003160 &            0.003194 &       0.002033 \\
     && 0.05 &   0.005232 &            0.004333 &       0.002864 \\
     && 0.08 &   0.008918 &            0.006315 &       0.004426 \\
     && 0.10 &   0.011887 &            0.005868 &       0.005576 \\ \hline \hline
         \multirow{16}{*}{I3\_60} & \multirow{4}{*}{BIM} & 0.03 &   0.002282 &            0.002345 &       0.002491 \\
    & & 0.05 &   0.003171 &            0.002995 &       0.002991 \\
    & & 0.08 &   0.004285 &            0.004185 &       0.003489 \\
    & & 0.10 &   0.005079 &            0.005460 &       0.003841 \\ \cline{2-6}
    & \multirow{4}{*}{FGSM} & 0.03 &   0.002055 &            0.002149 &       0.002446 \\
   & & 0.05 &   0.002551 &            0.002617 &       0.002898 \\
   & & 0.08 &   0.003173 &            0.003318 &       0.003361 \\
   & & 0.10 &   0.003508 &            0.003782 &       0.003578 \\ \cline{2-6}
     & \multirow{4}{*}{MIM} & 0.03 &   0.002270 &            0.002347 &       0.002496 \\
    & & 0.05 &   0.003161 &            0.003063 &       0.003006 \\
    & & 0.08 &   0.004354 &            0.004522 &       0.003566 \\
    & & 0.10 &   0.005245 &            0.005896 &       0.003998 \\ \cline{2-6}
     & \multirow{4}{*}{PGD} & 0.03 &   0.002282 &            0.002345 &       0.002491 \\
    & & 0.05 &   0.003171 &            0.002995 &       0.002991 \\
    & & 0.08 &   0.004285 &            0.004184 &       0.003489 \\
    & & 0.10 &   0.005079 &            0.005461 &       0.003841 \\
\hline \hline
    \multirow{16}{*}{I2\_2p5} & \multirow{4}{*}{BIM} & 0.03 &   0.001113 &            0.000268 &       0.002583 \\
    & & 0.05 &   0.002655 &            0.001029 &       0.002548 \\
    & & 0.08 &   0.005145 &            0.002399 &       0.002586 \\
    & & 0.10 &   0.006786 &            0.003317 &       0.002649 \\ \cline{2-6}
    & \multirow{4}{*}{FGSM} & 0.03 &   0.000601 &            0.000138 &       0.002582 \\
   & & 0.05 &   0.001314 &            0.000479 &       0.002541 \\
   & & 0.08 &   0.002288 &            0.000989 &       0.002551 \\
   & & 0.10 &   0.002818 &            0.001428 &       0.002583 \\ \cline{2-6}
    & \multirow{4}{*}{MIM} & 0.03 &   0.001048 &            0.000252 &       0.002583 \\
    & & 0.05 &   0.002540 &            0.001003 &       0.002549 \\
    & & 0.08 &   0.005054 &            0.002457 &       0.002593 \\
    & & 0.10 &   0.006742 &            0.003453 &       0.002663 \\ \cline{2-6}
    & \multirow{4}{*}{PGD} & 0.03 &   0.001113 &            0.000268 &       0.002583 \\
    & & 0.05 &   0.002655 &            0.001029 &       0.002548 \\
    & & 0.08 &   0.005144 &            0.002399 &       0.002586 \\
    & & 0.10 &   0.006785 &            0.003317 &       0.002649 \\ \hline
\end{tabular}
    \label{tab:rq3}
\end{table}

\fbox{\begin{minipage}{1.0\linewidth}
\textbf{Concluding Remarks for RQ3:} The defensive distillation mitigation method is more resilient against higher-order attacks, which are more difficult to detect.
\end{minipage}}

\section{Discussion}
\label{sec:discussion}

In this study, a comprehensive analysis of the mmWave beamforming prediction model's vulnerabilities and mitigations has been provided. The model's vulnerabilities are investigated for various adversarial attacks, i.e., FGSM, BIM, PGD, and MIM, while the mitigations for  adversarial training and defensive distillation are explored.  The results show that mmWave beamforming prediction models provide a satisfactory performance  without any adversarial attacks. On the other hand, the models are very sensitive to adversarial attacks, especially BIM. For example, as shown in Figure \ref{fig:rq1}, the MSE value can rise to 0.5 (for O1 scenario) and 0.20 (for I1\_2p5, and I3\_60 scenarios) under a heavy adversarial attack, i.e., $\epsilon=0.9$. According to Figure \ref{fig:O1_rq2}-\ref{fig:I1_2p5_rq2}, the MSE distribution of the model performance is not uniform under the adversarial attack. This is because those attacks add noise to the input and/or target features.  Figure \ref{fig:rq3} demonstrates the adversarial training and defensive distillation-based mitigation methods results on mmWave beamforming prediction as a summary. The defensive distillation mitigation method provides a better performance against higher-order attacks. 

Observations derived from the results of adversarial attacks on mmWave beamforming prediction models and the use of mitigation methods can be summarized as:\\
\emph{\textit{Observation 1}}: The mmWave beamforming prediction models are vulnerable to adversarial attacks.\\
\emph{\textit{Observation 2}}: BIM is the most successful
attack among those selected for study..\\
\emph{\textit{Observation 3}}: There is a strong negative correlation between attack power  \(\epsilon\) and the performance of DL-based mmWave beamforming prediction models.\\
\emph{\textit{Observation 4}}: The defensive distillation mitigation method is more resilient against higher-order attacks.\\

\section{Conclusion and Future Works}
\label{sec:conclusion}
This paper presents a DL security scheme for RF beamforming prediction models' vulnerabilities and their mitigation techniques by satisfying the following research questions:  (1) Can we generate malicious inputs for beamforming vector prediction models using FGSM, PGD, BIM, and MIM attacks in the complex domain?; (2)Is there any correlation between noise vector norm value (i.e., epsilon) and prediction performance with MSE metric?; and (3) What are the adversarial training based mitigation methods’ protection performance metric results with different epsilon values? The experiments were performed with the selected DeepMIMO scenarios, i.e., \textit{O1\_60, I1\_2p5, and I3\_60 ray-tracing} to investigate these questions.  The results confirm that the original DL-based beamforming model is significantly vulnerable to FGSM, PGD, BIM, and MIM attacks, especially BIM. The MSE value increases in all three scenarios under a heavy BIM adversarial attack (\(\epsilon\)= 0.9), i.e., 0.5 (for O1\_60 scenario) and 0.20 (for I1\_2p5 and I3\_60 scenarios). There is a high negative correlation between attack power  \(\epsilon\) and the performance of models, i.e., a high \(\epsilon\) increases as the model's performance dramatically decreases. On the other hand, the results show that the proposed mitigation methods, i.e., the iterative adversarial training and defensive distillation approaches, successfully increase the RF beamforming prediction performance and create more accurate predictions. The overall results prove that the proposed framework can enhance the DL-based beamforming model performance. As future work, the research team plans to investigate other AI-based solutions used in physical and media access layers of 6G networks, i.e.,  channel coding, synchronization, positioning, channel estimations, symbol detection, resource allocation, and scheduling, and their cybersecurity risks.

%

\bibliographystyle{plain}
\bibliography{refs}

\begin{thebibliography}{10}

\bibitem{I1_s}
Deep{MIMO}, {'I1' scenario}.
\newblock \url{https://deepmimo.net/scenarios/i1-scenario/}.
\newblock Accessed: 2021-09-30.

\bibitem{I3_s}
Deep{MIMO}, {'I3' scenario}.
\newblock \url{https://deepmimo.net/scenarios/i3-scenario/}.
\newblock Accessed: 2021-09-30.

\bibitem{O1_s}
Deep{MIMO}, {'O1' scenario}.
\newblock \url{https://deepmimo.net/scenarios/o1-scenario/}.
\newblock Accessed: 2021-09-30.

\bibitem{RemCom}
Remcom, {Wireless InSite}.
\newblock \url{http://www.remcom.com/wireless-insite}.
\newblock Accessed: 2021-09-30.

\bibitem{ali20206g}
Samad Ali, Walid Saad, Nandana Rajatheva, Kapseok Chang, Daniel Steinbach,
  Benjamin Sliwa, Christian Wietfeld, Kai Mei, Hamid Shiri, Hans-Jürgen
  Zepernick, Thi My~Chinh Chu, Ijaz Ahmad, Jyrki Huusko, Jaakko Suutala,
  Shubhangi Bhadauria, Vimal Bhatia, Rangeet Mitra, Saidhiraj Amuru, Robert
  Abbas, Baohua Shao, Michele Capobianco, Guanghui Yu, Maelick Claes, Teemu
  Karvonen, Mingzhe Chen, Maksym Girnyk, and Hassan Malik.
\newblock 6{G} white paper on machine learning in wireless communication
  networks, 2020.

\bibitem{alkhateeb2019deepmimo}
Ahmed Alkhateeb.
\newblock Deep{MIMO}: A generic deep learning dataset for millimeter wave and
  massive {MIMO} applications.
\newblock {\em arXiv preprint arXiv:1902.06435}, 2019.

\bibitem{2020arXiv200702617A}
Maksym {Andriushchenko} and Nicolas {Flammarion}.
\newblock {Understanding and Improving Fast Adversarial Training}.
\newblock {\em arXiv e-prints}, page arXiv:2007.02617, July 2020.

\bibitem{2021arXiv210201356B}
Tao {Bai}, Jinqi {Luo}, Jun {Zhao}, Bihan {Wen}, and Qian {Wang}.
\newblock {Recent Advances in Adversarial Training for Adversarial Robustness}.
\newblock {\em arXiv e-prints}, page arXiv:2102.01356, February 2021.

\bibitem{9527756}
Evren Catak, Ferhat~Ozgur Catak, and Arild Moldsvor.
\newblock Adversarial machine learning security problems for 6{G}: mmwave beam
  prediction use-case.
\newblock In {\em 2021 IEEE International Black Sea Conference on
  Communications and Networking (BlackSeaCom)}, pages 1--6, 2021.

\bibitem{CATAK2017184}
Evren Catak and Lutfiye Durak-Ata.
\newblock Adaptive filterbank-based multi-carrier waveform design for flexible
  data rates.
\newblock {\em Computers \& Electrical Engineering}, 61:184--194, 2017.

\bibitem{catak2021security}
Ferhat~Ozgur Catak, Murat Kuzlu, Evren Catak, Umit Cali, and Devrim Unal.
\newblock Security concerns on machine learning solutions for 6{G} networks in
  mm{W}ave beam prediction.
\newblock {\em Physical Communication}, page 101626, 2022.

\bibitem{dang2020should}
Shuping Dang, Osama Amin, Basem Shihada, and Mohamed-Slim Alouini.
\newblock What should 6{G} be?
\newblock {\em Nature Electronics}, 3(1):20--29, 2020.

\bibitem{de2021survey}
Chamitha De~Alwis, Anshuman Kalla, Quoc-Viet Pham, Pardeep Kumar, Kapal Dev,
  Won-Joo Hwang, and Madhusanka Liyanage.
\newblock Survey on 6{G} frontiers: Trends, applications, requirements,
  technologies and future research.
\newblock {\em IEEE Open Journal of the Communications Society}, 2:836--886,
  2021.

\bibitem{9206115}
Jun Du, Chunxiao Jiang, Jian Wang, Yong Ren, and Merouane Debbah.
\newblock Machine learning for 6{G} wireless networks: Carrying forward
  enhanced bandwidth, massive access, and ultrareliable/low-latency service.
\newblock {\em IEEE Vehicular Technology Magazine}, 15(4):122--134, 2020.

\bibitem{2021arXiv210204150F}
Omer {Faruk Tuna}, Ferhat {Ozgur Catak}, and M.~{Taner Eskil}.
\newblock {Exploiting epistemic uncertainty of the deep learning models to
  generate adversarial samples}.
\newblock {\em arXiv e-prints}, page arXiv:2102.04150, February 2021.

\bibitem{fostiropoulosrobust}
Iordanis Fostiropoulos, Basel Shbita, and Myrl Marmarelis.
\newblock Robust defense against lp-norm-based attacks by learning robust
  representations.

\bibitem{giordani2020toward}
Marco Giordani, Michele Polese, Marco Mezzavilla, Sundeep Rangan, and Michele
  Zorzi.
\newblock Toward 6{G} networks: Use cases and technologies.
\newblock {\em IEEE Communications Magazine}, 58(3):55--61, 2020.

\bibitem{9023459}
Guan Gui, Miao Liu, Fengxiao Tang, Nei Kato, and Fumiyuki Adachi.
\newblock 6{G}: Opening new horizons for integration of comfort, security, and
  intelligence.
\newblock {\em IEEE Wireless Communications}, 27(5):126--132, 2020.

\bibitem{hinton2015distilling}
Geoffrey Hinton, Oriol Vinyals, and Jeff Dean.
\newblock Distilling the knowledge in a neural network, 2015.

\bibitem{jiang2021project}
Yan Jiang, Guisheng Yin, Ye~Yuan, and Qingan Da.
\newblock Project gradient descent adversarial attack against multisource
  remote sensing image scene classification.
\newblock {\em Security and Communication Networks}, 2021, 2021.

\bibitem{9163104}
Latif~U. Khan, Ibrar Yaqoob, Muhammad Imran, Zhu Han, and Choong~Seon Hong.
\newblock 6{G} wireless systems: A vision, architectural elements, and future
  directions.
\newblock {\em IEEE Access}, 8:147029--147044, 2020.

\bibitem{kuzlu2021role}
Murat Kuzlu, Corinne Fair, and Ozgur Guler.
\newblock Role of artificial intelligence in the internet of things ({IoT})
  cybersecurity.
\newblock {\em Discover Internet of Things}, 1(1):1--14, 2021.

\bibitem{8403769}
Marc Lichtman, Raghunandan Rao, Vuk Marojevic, Jeffrey Reed, and Roger~Piqueras
  Jover.
\newblock 5{G} {NR} jamming, spoofing, and sniffing: Threat assessment and
  mitigation.
\newblock In {\em 2018 IEEE International Conference on Communications
  Workshops (ICC Workshops)}, pages 1--6, 2018.

\bibitem{9259112}
Yun Lin, Haojun Zhao, Xuefei Ma, Ya~Tu, and Meiyu Wang.
\newblock Adversarial attacks in modulation recognition with convolutional
  neural networks.
\newblock {\em IEEE Transactions on Reliability}, 70(1):389--401, 2021.

\bibitem{liu20185g}
Guangyi Liu, Yuhong Huang, Fei Wang, Jianjun Liu, and Qixing Wang.
\newblock 5{G} features from operation perspective and fundamental performance
  validation by field trial.
\newblock {\em China Communications}, 15(11):33--50, 2018.

\bibitem{michels2019vulnerability}
Felix Michels, Tobias Uelwer, Eric Upschulte, and Stefan Harmeling.
\newblock On the vulnerability of capsule networks to adversarial attacks.
\newblock {\em arXiv preprint arXiv:1906.03612}, 2019.

\bibitem{ozpoyraz2022deep}
Burak Ozpoyraz, A.~Tugberk Dogukan, Yarkin Gevez, Ufuk Altun, and Ertugrul
  Basar.
\newblock Deep learning-aided 6{G} wireless networks: A comprehensive survey of
  revolutionary phy architectures, 2022.

\bibitem{papernot2016distillation}
Nicolas Papernot, Patrick McDaniel, Xi~Wu, Somesh Jha, and Ananthram Swami.
\newblock Distillation as a defense to adversarial perturbations against deep
  neural networks, 2016.

\bibitem{porambage20216g}
Pawani Porambage, G{\"u}rkan G{\"u}r, Diana Pamela~Moya Osorio, Madhusanka
  Liyanage, and Mika Ylianttila.
\newblock 6{G} security challenges and potential solutions.
\newblock In {\em Proc. IEEE Joint Eur. Conf. Netw. Commun.(EuCNC) 6{G}
  Summit}, pages 1--6, 2021.

\bibitem{saad2019vision}
Walid Saad, Mehdi Bennis, and Mingzhe Chen.
\newblock A vision of 6{G} wireless systems: Applications, trends,
  technologies, and open research problems.
\newblock {\em IEEE network}, 34(3):134--142, 2019.

\bibitem{sheth2020taxonomy}
Karan Sheth, Keyur Patel, Het Shah, Sudeep Tanwar, Rajesh Gupta, and Neeraj
  Kumar.
\newblock A taxonomy of {AI} techniques for 6{G} communication networks.
\newblock {\em Computer Communications}, 161:279--303, 2020.

\bibitem{siriwardhana2021ai}
Yushan Siriwardhana, Pawani Porambage, Madhusanka Liyanage, and Mika
  Ylianttila.
\newblock {AI} and 6{G} security: Opportunities and challenges.
\newblock In {\em Proc. IEEE Joint Eur. Conf. Netw. Commun.(EuCNC) 6{G}
  Summit}, pages 1--6, 2021.

\bibitem{vardhan2021ensemble}
Raj Vardhan.
\newblock {\em An Ensemble Approach for Explanation-based Adversarial
  Detection}.
\newblock PhD thesis, 2021.

\bibitem{9237460}
Helin Yang, Arokiaswami Alphones, Zehui Xiong, Dusit Niyato, Jun Zhao, and
  Kaishun Wu.
\newblock Artificial-intelligence-enabled intelligent 6{G} networks.
\newblock {\em IEEE Network}, 34(6):272--280, 2020.

\bibitem{zhang20196g}
Zhengquan Zhang, Yue Xiao, Zheng Ma, Ming Xiao, Zhiguo Ding, Xianfu Lei,
  George~K Karagiannidis, and Pingzhi Fan.
\newblock 6{G} wireless networks: Vision, requirements, architecture, and key
  technologies.
\newblock {\em IEEE Vehicular Technology Magazine}, 14(3):28--41, 2019.

\bibitem{zheng2021potential}
Zunxin Zheng, Linmei Wang, Fumin Zhu, and Ling Liu.
\newblock Potential technologies and applications based on deep learning in the
  6{G} networks.
\newblock {\em Computers \& Electrical Engineering}, 95:107373, 2021.

\end{thebibliography}

\end{document}